\documentclass[12pt]{spieman}  
\usepackage[]{aas_macros}  
\usepackage{amsmath,amsfonts,amssymb}
\usepackage{graphicx}
\usepackage{setspace}
\usepackage{tocloft}
\usepackage{wrapfig}
\usepackage[symbol]{footmisc}

\def\b1{\boldsymbol{1}}

\def\ba{\boldsymbol{a}}

\def\bb{\boldsymbol{b}}

\def\bc{\boldsymbol{c}}

\def\bC{\boldsymbol{C}}
\newcommand{\rd}{\mathrm{d}} 
 
\def\bD{\boldsymbol{D}}

\def\bE{\boldsymbol{E}}

\def\bF{\boldsymbol{F}}
\def\bff{\boldsymbol{f}}

\def\bh{\boldsymbol{h}}

\def\bi{\boldsymbol{i}}

\def\rm{\mathrm{m}}

\def\bn{\boldsymbol{n}}

\def\bp{\boldsymbol{p}}

\def\br{\boldsymbol{r}}

\def\rt{\mathrm{t}}

\def\rT{\mathrm{T}}

\def\bu{\boldsymbol{u}}
\def\bv{\boldsymbol{v}}

\def\bx{\boldsymbol{x}}
\def\by{\boldsymbol{y}}

\def\bz{\boldsymbol{z}}

\def\balpha{\boldsymbol{\alpha}}

\def\brho{\boldsymbol{\rho}}

\title{A Laboratory Method for Measuring the Cross-Polarization in High-Contrast Imaging}

\author[$\dagger$]{Richard A. Frazin}
\affil[$\dagger$]{Dept. of Climate and Space Sciences, University of Michigan, Ann Arbor, MI 48109}

\cftpagenumbersoff{figure}
\cftpagenumbersoff{table} 
\begin{document} 
\maketitle


\begin{abstract}

Electric Field Conjugation (EFC) and related techniques have proven to be effective for stellar coronagraphy, but existing methods are not suited for treating small polarization effects, referred to here as cross polarization, resulting from propagation through the optical system. EFC utilizes a deformable mirror (DM) to nullify the electric fields from the host star within a \emph{dark hole} region in the image plane, allowing for the detection of much fainter planets.

This article presents realistic numerical simulations of a laboratory method for measuring the cross-polarization electric field in a stellar coronagraph. This novel method uses linear polarizers to isolate the cross polarization component and probing procedures similar to those in existing EFC methods to measure the electric field corresponding to cross polarization. The proposed nonlinear probing scheme is specifically designed to address polarizer leakage. The method is not a generalization of existing EFC procedures in order to null the cross polarization field; rather, it is a non-iterative laboratory calibration procedure with applications in instrument characterization and validation of numerical models. However, lab-validated numerical models of the cross-polarization are directly applicable to on-sky observations.

The proposed experimental method is not significantly more demanding than existing testbed procedures, requiring little additional equipment. Yet, the simulation results demonstrate highly accurate estimates of the electric field corresponding to the cross polarization in a dark hole. These encouraging results suggest that laboratory implementation is viable.
Simulations are performed for a randomly aberrated Lyot coronagraph with a dark hole size of $\sim 4 \lambda/D \times 4 \lambda/D$ in the image plane with a contrast of $\sim 10^{-10}$ in contrast (i.e., the planet-to-star intensity ratio) units.
The cross-polarization electric fields accurately estimated by the method correspond to intensities of $\sim 10^{-11}$ in contrast units.

\end{abstract}


{\noindent \footnotesize\textbf{$\dagger$}Richard Frazin,  \linkable{rfrazin\emph{\_at\_}umich.edu} }

\section{Introduction}\label{sec: intro}

Direct imaging of exoplanets is among NASA's top priorities in future missions, such as the Large UV/Optical/Infrared Surveyor (LUVOIR).
The principal challenge of direct imaging is the fact that exoplanets are located in close angular proximity to their vastly brighter host stars. 
Quantitatively, this means that the telescopic optical system must be able to achieve a contrast ratio, i.e., planet-to-star brightness ratio, of less than $10^{-8}$, or perhaps even $10^{-11}$, depending on the target.\cite{LUVOIR_model_JATIS22}
Current designs to meet this daunting requirement feature stellar coronagraphs, which extinguish on-axis light but allow slightly off-axis beams to pass through relatively unimpeded.\cite{Cavarroc_IdealCoronagraph06}
Even if the stellar coronagraphs had perfect optical surfaces, diffraction alone would put the contrast levels orders of magnitude above the aforementioned requirements.
Real surfaces have aberrations at high and low spatial frequencies that further degrade the contrast.\cite{Krist_End2End_Roman_JATIS23}
Achieving high contrast requires active wavefront control strategies, also known as \emph{adaptive optics}.
In the space-based context, the most prominent of these is a family of methods that fall under the term \emph{electric field conjugation (EFC)}.
EFC procedures use one or two deformable mirrors (DMs) to modulate the intensity measured in the image plane through alternating sensing and control steps.
The end result of EFC procedures is a region of the image plane called a  \emph{dark hole}  in which the starlight is suppressed to a high contrast level.\cite{GiveonKern_EFC11,Kasdin_EFC16b}
In test bed settings, such procedures yield dark holes with roughly $10^{-9}$ of the brightness of the laser source, which plays the role of a star.\cite{Belikov_LabDemo_SPIE22, Seo_JPLhiContrastResult_JATIS19}

EFC procedures are based on the fact that the starlight present at some level throughout the image plane is coherent, which allows the DM to modulate the intensity and create the dark hole, which is a region of destructive interference.
Any exoplanet light present will be slightly off-axis and incoherent with the starlight.
This incoherence is the key to detection of faint exoplanets, since the center of the exoplanet's image will not be modulated significantly by the DM.\cite{Bottom_CDI_MNRAS17}
Thus, the portion of the light that is not modulated is ascribed to planetary emission.
One limitation to the contrast that can be achieved via such procedures is imposed by polarization effects, which can be confounded with incoherence at very high contrast levels, as this article explains in detail later.
Baudoz \emph{et al.} argued that polarization aberration had an effect of $\sim10^{-8}$ on the TDH2 bench.\cite{Baudoz_Goos-Hanchen_Imbert-Fedorov}
Further, polarization-dependent aberrations were part of a comprehensive study on the effect of aberrations the Roman Space Telescope Coronagraph, showing that $10^{-9}$ or better contrast is within reach by Krist \emph{et al.}\cite{Krist_End2End_Roman_JATIS23}

\subsection{Summary of Article Content}\label{sec: what to expect}

This article presents simulations of a concept for a relatively simple, yet highly sensitive, laboratory method for measuring the electric fields corresponding to the contribution of the cross polarization to the total intensity in a dark hole.  
Its most likely uses are laboratory calibration of cross polarization and validation of cross polarization models, which is timely given the increasing importance of digital twin concepts in high contrast imaging.\cite{Haffert_DigiTwin2024} 
At the simplest level, the proposed measurements could validate models of basic descriptive statistics (e.g., mean and variance) of the cross-polarization intensity.
More sophisticated would be validating AI-based models that map measurements of the dominant field made in the course of classical EFC procedures to the cross-polarization field (``cross field'' for short).
Lab-validated models of the cross field would potentially improve the achievable contrast in on-sky observations because polarization effects could be taken into account in the estimate of the planetary signal or possibly even nulled, but these aspects are not the subject of this article.
  
While this method uses probing and concepts that are similar to those in the EFC sensing step, it is critical to understand that this method is not a generalization of EFC.  This method only measures the cross field, and it is not iterative.  
Assuming access to a high-contrast testbed with a coronagraph and working EFC procedures to create a dark hole, the additional lab equipment required includes a 100 mW supercontinuum laser with a continuous wave (CW) mode and two linear polarizers with less than $10^{-5}$ leakage of the orthogonal intensity.
The proposed procedure consists of the following steps, which are explained in more detail in the course of this article:
\begin{enumerate}
\item Without the linear polarizers in place, apply EFC procedures to make a dark hole.  Take note of the DM command corresponding to the dark hole, $\bc_0$.
\item Insert the two linear polarizers.  One goes just after the entrance pupil and the other goes just before the detector.  They are arranged so that if there were no polarizing effects in the optical system, all of the light would be extinguished, leakage from the polarizers notwithstanding.  This arrangement isolates the cross polarization arriving at the detector.
Note that isolating the cross polarization in this way comes at the expense of extinguishing any planetary light, which does not conflict with the objective of measuring the cross field.  (Indeed, one is free to remove the polarizers after making the measurements.)
\item Apply the probing procedures detailed in this article to measure the intensities needed to feed the regression method.
\item Apply the regression method detailed in this article to estimate the electric field associated with the cross polarization in the dark hole.
\end{enumerate}

Sec.~\ref{sec: cross pol examples} provides simulation examples of cross-polarization arising in simple focusing optical systems.
Sec.~\ref{sec: Intensity with CP} elucidates the relationship between coherence, polarization and intensity, as applied in this article.
Sec.~\ref{sec: models} presents the matrix-based coronagraph models underpinning the numerical methods presented here and introduces the dark hole.
Sec.~\ref{sec: Probing} details the probing and analysis procedures for estimating the cross fields while accounting for leakage of orthogonal intensity from the linear polarizers.
Sec.~\ref{sec: simulations} describes the numerical simulations of the proposed method and the resulting estimates of the cross field.
Finally, Sec.~\ref{sec: Conclusions}  reviews key points of this work and discusses various practical aspects of carrying out the lab experiment, namely laser fluctuations, stray light from the polarizers and wavefront error introduced by the polarizers.  In addition, it outlines how one might make use of a lab-validated model of the cross field on-sky.

\section{Examples of Cross-Polarization in Simple Focusing Systems}\label{sec: cross pol examples} 

\begin{figure}[t]
\includegraphics[height=3.5cm]{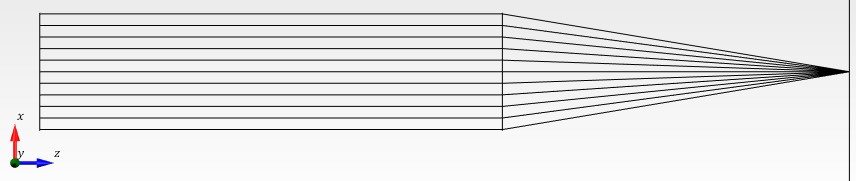}
\caption{\label{fig: beam to focus}
A simple $f$/\#~=~3 optical system that brings rays to a focus with an idealized thin lens. } 
\end{figure}

It is generally appreciated that reflections can have polarization effects, which are calculated in a straightforward manner with the Fresnel coefficients for a plane wave reflecting from a planar surface.\cite{Born&Wolf}
Less well recognized is that the finite $f$/\# (``f-number") of a focusing optical system can convert some amount of one linear polarization into orthogonal linear polarization.
We will refer to the generation of the orthogonal polarization via nominally unpolarized powered or planar optics as \emph{cross-polarization}.
To understand cross-polarization with powered optics, consider a collimated beam with diameter $D$ propagating along the $+z$ axis and polarized in the $x$ direction with field strength $\sqrt{I}$, where $I$ is the intensity of the collimated beam.
The light is then refracted by a thin lens with focal length $f$ and brought to a focus, giving this system $f$/\# $=f/D$.
This is depicted in Fig.~\ref{fig: beam to focus}, which shows the $x-z$ plane of a simple optical system that has cylindrical symmetry about the $z$ axis. 

\begin{figure}[] 
	\hspace{-2mm}
	\begin{tabular}{l l l}
		\includegraphics[height=4.5cm]{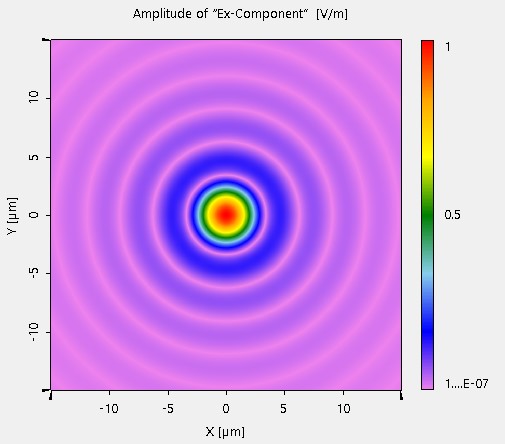} &
		\includegraphics[height=4.5cm]{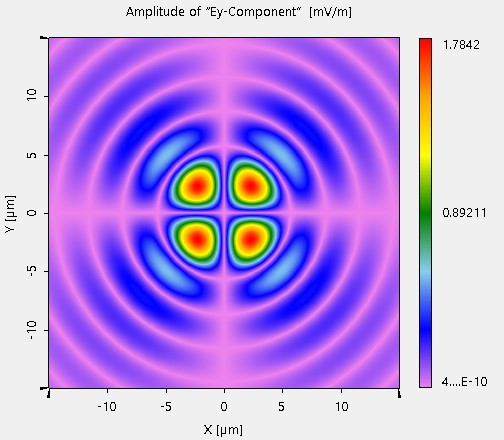} &
		\includegraphics[height=4.5cm]{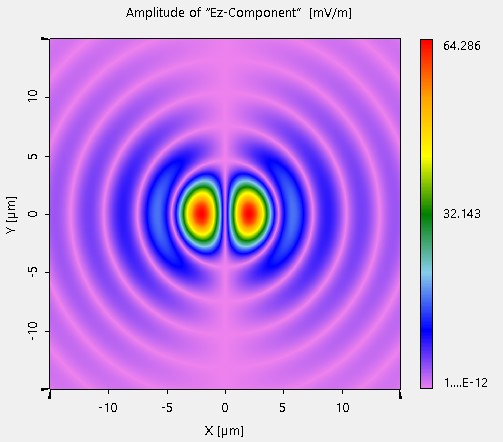} \\
		\includegraphics[height=4.5cm]{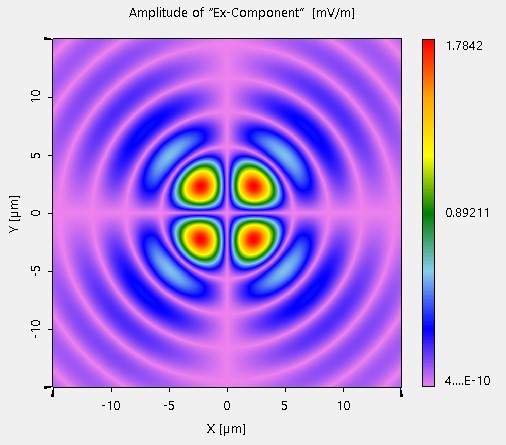} &
		\includegraphics[height=4.5cm]{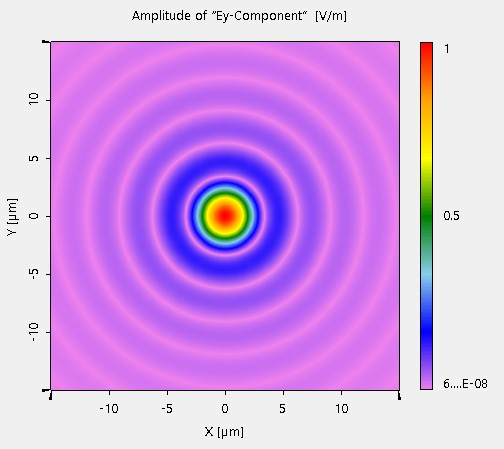} &
		\includegraphics[height=4.5cm]{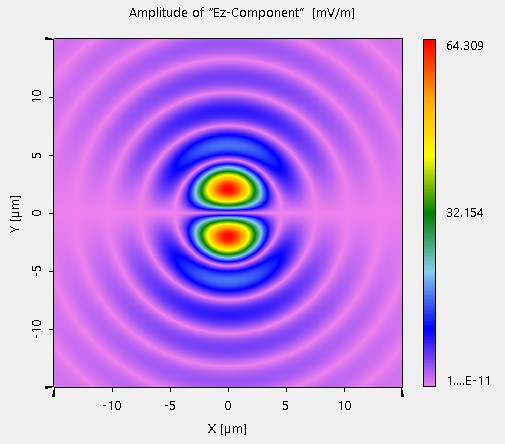} 
	\end{tabular}
	\caption{\label{fig: fields simple}
		Focal plane electric fields corresponding to the $f$/\#~=~3 optical system depicted in Fig.~\ref{fig: beam to focus}
		[diameter $D=10\,$cm, focal length $f=30\,$cm, wavelength $\lambda=1\,\mu$m].
		The top row shows $|E_x|$ (peak~=~1~V/m), $|E_y|$ (peak~=~1.8~mV/m), and $|E_z|$ (peak~=~64~mV/m), respectively, when the input polarization is in the $x$ direction.
		The bottom row shows $|E_x|$ (peak~=~1.8~mV/m), $|E_y|$ (peak~=~1~V/m), and $|E_z|$ (peak~=~64~mV/m), respectively, when the input polarization is in the $y$ direction.
		The dominant polarization is shown in the upper right and middle bottom; the others are cross-polarization.
		All field units are V/m and are normalized by a factor that makes the maximum of $|E_x|$ in the top row 1 V/m.
		The color scale in all images is linear.
		Note the symmetry in that $|E_x|$ in the top row is the same as $|E_y|$ in the bottom row and that $|E_x|$ in the bottom row is the same as $|E_y|$ in the top row.  
	} 
\end{figure} 

\begin{figure}[]
	\hspace{-3mm}
	\begin{tabular}{l l}
		\includegraphics[height=7.3cm]{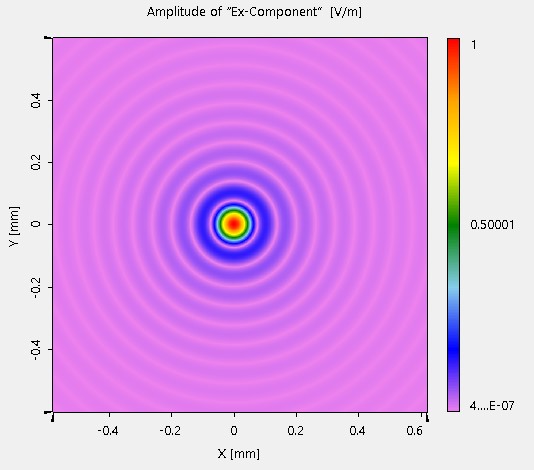} & \hspace{-6mm}
		\includegraphics[height=7.3cm]{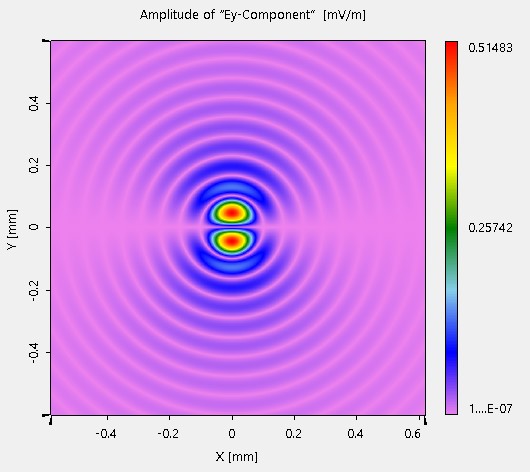} \\
		\includegraphics[height=7.3cm]{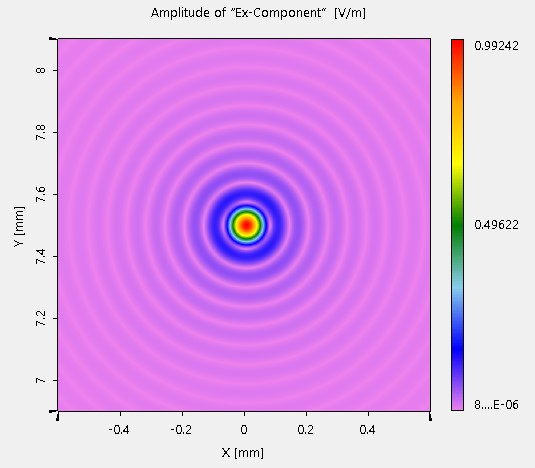} & \hspace{-6mm}
		\includegraphics[height=7.3cm]{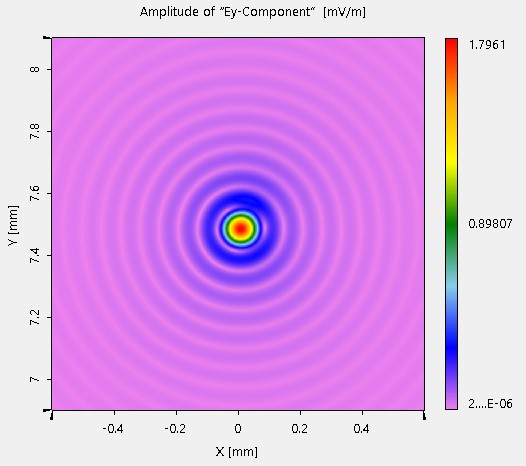} 
	\end{tabular}
	\caption{\label{fig: fields OAP}
		Focal plane electric fields corresponding to a collimated beam focused by a $f$/\#~=~62.5 OAP with an off-axis reflection angle of $20^\circ$.  
		See text for details.
		The units of the linear color scale are V/m in the left column and mV/m in the right column.
		The values are normalized so that the peak of the upper left panel is 1~V/m.
		The upper row corresponds to an on-axis input beam (focal point at the origin), while the lower row shows results for an off-axis input beam with a focal point at $(0, \, 7.5 \, \mathrm{mm})$.
		The input beams in both cases are $x$-polarized.
		The two images on the left are dominant polarization and the two on the right are cross-polarization.
		The upper-left image shows $|E_x|$ (peak~=$\, 1\,$V/m)  and the upper-right image shows $|E_y|$ (peak~=$ \, 0.5\,$mV/m).
		The lower-left image shows $|E_x|$ (peak~=$\, 0.992\,$V/m), and the lower-right image shows $|E_y|$ (peak~=$\,1.8\,$mV/m).
		Note the differences shown in the images on the right.
		There is no need to show results for $y$-polarized input beams, since the results are symmetrical as they are in Fig.~\ref{fig: fields simple}.
	} 
\end{figure}

While none of the rays in the $x-z$ plane shown in the diagram will result in $y$ polarization (this is also true for $x$-polarized rays in the $y-z$ plane), the rays in this plane do result in a $z$ cross-polarization component:
Consider the ray in the diagram that hits the lens a distance $x_0$ from the $z$ axis.
The refraction angle of this ray, $\theta_0$ is specified by $\tan\theta_0 = x_0/f$, and for a large $f$/\# we approximate $\theta_0 \approx x_0/f$.
To find the electric field vector of this ray just after it has been refracted by the lens, we must rotate the electric field vector of the collimated beam, $\sqrt{I}(1, \, 0, \, 0)$, about the $y$ axis (because the ray is in the $x-z$ plane) by the angle $\theta_0$.
The direction vector of the refracted electric field corresponding to this ray is $\hat{\bu} = (\cos \theta_0, \, 0, \, \sin \theta_0)$, and the electric field vector for this ray is simply $ \hat{\bu} \sqrt{I}$.

The trigonometry is trickier when we treat rays that are not in $x-z$ or $y-z$ planes.
If we consider an $x$-polarized ray in the collimated beam that hits the lens at a position $(x_0,\, y_0, \, 0)$ (taking $z=0$ to be the plane of the lens), this ray is refracted by angle of $\theta_0$, where $\tan \theta_0 = \rho_0 / f$ and $ \rho_0 = \sqrt{x_0^2 + y_0^2}$.
The direction of the refracted electric field is obtained by rotating the vector $(1, \,0,\,0)$ by the angle $\theta_0$ about the axis defined by the direction vector $(-y_0,\, x_0, \, 0)/ \rho_0$.  
Defining $\cos \psi_0 = x_0/\rho_0$ and $\sin \psi_0 = y_0/\rho_0$ the resulting unit vector in the direction of the electric field of the refracted ray is given by: 
[\emph{https:}en.wikipedia.org/wiki/Rotation\_matrix]
\begin{equation} \left(
\begin{array}{l}
\cos \theta_0 + \sin^2\psi_0(1-\cos \theta_0) \\  -\sin\psi_0 \cos\psi_0(1-\cos\theta_0), \\ -\cos\psi_0 \sin \theta_0 
\end{array} \right) \, .
\label{eq: Ex thin lens}
\end{equation}
On the other hand, if the collimated beam is instead $y$-polarized, the unit vector in the direction of the electric field of the refracted ray is given by:
\begin{equation} \left(
\begin{array}{l}
-\sin\psi_0 \cos\psi_0(1-\cos\theta_0), \\ \cos \theta_0 + \cos^2\psi_0(1-\cos \theta_0)  \\ -\sin\psi_0 \sin \theta_0 
\end{array} \right) \, .
\label{eq: Ey thin lens}
\end{equation}
In Eq.~\eqref{eq: Ex thin lens}, since the collimated beam is assumed to be $x$ polarized, cross-polarization components are the $y$ and $z$ components.   The $x$ component is called the \emph{dominant component}.
Similarly, Eq.~\eqref{eq: Ey thin lens}, the dominant component is the $y$ component and the $x$ and $z$ components are the cross-polarization components.
For the large $f$/\# systems of interest, the most important term in the cross-polarization components shown in Eqs.~\eqref{eq: Ex thin lens} and~\eqref{eq: Ey thin lens} is $\sin\theta_0 \approx \rho_0/f \leq D/f$, which means the cross-polarization is essentially proportional to the $f$/\#.

\subsection{Simulation Examples}\label{sec: Cross Pol Simulations}

The first example we present in this section corresponds to the optical system depicted in Fig.~\ref{fig: beam to focus}, which is a $f$/\#~=~3 system with a numerically ideal lens.
The second example is a realistic simulation of an off-axis parabola (OAP) with a much larger $f$/\#.

In the case of Fig.~\ref{fig: beam to focus}, ``ideal'' only means that the focusing is aberration-free (despite the extreme 
$f$/\#).
Although such a small $f$/\# would be useful in confocal microscopy, it is not useful for telescopes; however, it provides a helpful and relatively simple illustration of the cross-polarization phenomenon.  
Fig.~\ref{fig: fields simple} shows the corresponding magnitudes $|\mathrm{fields}|$ in the focal plane, with the top row corresponding to $x$-polarization in the input beam, and $y$-polarization in the bottom row.  
It is important to notice that $|E_x|$ in the top row is the same as $|E_y|$ in the bottom row; these are the \emph{dominant polarizations} defined in Sec.~\ref{sec: Intensity with CP}.
Similarly,  $|E_y|$ in the top row is the same as $|E_x|$ in the bottom row; these are the \emph{cross-polarizations} also defined in Sec.~\ref{sec: Intensity with CP}.
The analytical developments to follow will take advantage of this symmetry.

Fig.~\ref{fig: fields OAP} shows simulations of $|E_x|$ and $|E_y|$ for a $x$-polarized collimated beam focused by an OAP.  
The diameter of the collimated input beam is 8~mm, the wavelength of the light is $1 \, \mu$m, and the OAP's  off-axis reflection angle is 20$^\circ$ with a focal length of 500~mm, giving the system $f$/\#~=~62.5.
The OAP material is silver with a thin glass coating.
The simulation is designed so that an on-axis collimated input beam propagates in the $+z$ direction and the optical axis remains in the $x-z$ plane after the $20^\circ$ reflection from the OAP.  
The upper pair of images in Fig.~\ref{fig: fields OAP} corresponds to the on-axis input beam.
Here, ``on-axis'' means that the initial propagation direction is purely in the $+z$ direction.
The results for $y-$polarized input beams are not shown because they are perfectly symmetrical, just as they are in Fig.~\ref{fig: fields simple}.    
Unlike Fig.~\ref{fig: fields simple}, the cross-field ($E_y$ for $x$-polarized input) does not split into four quadrants, instead, it is split into 2 regions; this is due to the off-axis reflection breaking symmetry about the $x$-axis.

The lower pair of images corresponds to an off-axis input beam.  For this beam, the initial propagation direction is $\hat{\by} \sin\eta + \hat{\bz}\cos \eta  $, where $\eta = - 120 \, \lambda/D = - 0.86^\circ$, which is why these images are centered at $(0, \, 7.5\, \mathrm{mm})$, rather than the origin of the detector plane.  While the lower left image is similar to the upper left image (minor differences can be seen), the cross field shown in the lower right image is rather different than in the upper right image, including a peak value of $|E_y|$ more than three times greater.  
Additionally, while the upper right image exhibits a bimodal behavior, the lower right does not, nor does it exhibit the same symmetry about the $x$-axis.
The small difference in the peak values in the images on the left is due to the angular dependence of the Fresnel reflection coefficients applied at the OAP surface.
Again, the results for $y-$polarized input beams are not shown because they are nearly symmetrical.

\section{Calculation of Intensities Including Cross-Polarization}\label{sec: Intensity with CP}

This section provides mathematical derivation of intensity that includes cross-polarization.
Consider an unpolarized beam of starlight at the entrance pupil of an imaging system.  We will confine this discussion to quasi-monochromatic light at the wavelength $\lambda$.  
The unpolarized light can be divided into two orthogonal linear polarization components, $A$ and $B$.  
The direction of $A$ in the entrance pupil is chosen so that $A$-polarized light has its dominant polarization in the $x$ direction in the detector plane where an image is formed.  
Similarly, the $B$ direction is chosen so that $B$-polarized light has its dominant component in the $y$ direction of the imaging plane.  
In other words, if we were to ignore the polarization effects of the imaging system, $A$-polarized light would result only in $x$-polarized light at the detector, and $B$-polarized light would result only in $y$-polarized light at the detector.
In the examples given above, the $y$-axis does not change when the beam encounters the focusing optic, so the $A$ direction is the $x$-axis and the $B$ direction is along the $y$-axis.  
Of course, most astronomical imaging systems that have coordinate system transformations need to be taken into account in order to determine what the$A$ and $B$ directions happen to be.

The time-dependent electric field vectors in the entrance pupil corresponding to the $A$-polarized and $B$-polarized light are given by the functions
$\hat{\ba} \, a(t) \exp(j2 \pi \nu t)$ and $\hat{\bb}\, b(t) \exp(j 2 \pi \nu t) $, where $\nu = c/\lambda$ ($c$ being the speed of light), $\hat{\ba}$ and $\hat{\bb}$ are orthogonal unit vectors in the $A$ and $B$ directions.
$a(t)$ and $b(t)$ are complex-valued stochastic processes, called \emph{envelope functions} that rapidly modulate the harmonic fields.\cite{Collett93}
The envelope functions $a(t)$ and $b(t)$ satisfy the following conditions:
\begin{align}
& \mathcal{P}[a(t),b(t')]  = \mathcal{P}[a(t)] \mathcal{P}[b(t')] \label{eq: stat indep} \\
& \overline{a(t)}  = \overline{b(t)}  = 0. \label{eq: zero mean envolopes} \\
& \overline{a(t)a^*(t)} = \overline{b(t)b^*(t)} = 1 \label{eq: unity variance} \\
& \overline{a(t)b(t)}  = \overline{a(t)b^*(t)}  = \overline{a^*(t)b(t)} = 0  \label{eq: incoherent envolopes}
 \, ,
\end{align}
where the $\mathcal{P}$ represents probability, the superscript $^*$ indicates complex conjugation and the overbar indicates a time-average operator, which for the purposes of this article is equivalent to taking the mean of a stochastic process.   
In reality, integration over a finite amount of time is needed for Eqs.~\eqref{eq: zero mean envolopes} through~\eqref{eq: incoherent envolopes} to be effectively realized.
We assume that the required integration times are much less than any currently possible detector frame rate (say, $10^{-4}\,$s).
Eq.~\eqref{eq: stat indep} states that the processes $a(t)$ and $b(t)$ are statistically independent.
Eq.~\eqref{eq: zero mean envolopes} states that envelope functions are zero-mean, and Eq.~\eqref{eq: incoherent envolopes} states that the envelope functions are \emph{incoherent}.  
In fact, Eq.~\eqref{eq: incoherent envolopes} follows from Eqs.~\eqref{eq: stat indep} and Eq.~\eqref{eq: zero mean envolopes}.
Eq.~\eqref{eq: unity variance} states that the envelope functions have a variance of unity.

It is clear that while $A$-polarized light mostly results in $x$-polarized light at the detector, there may also be some $y$ and $z$ polarized light as well.
An analogous statement applies to $B$-polarized light.
Assuming that the $z$ direction is normal to the detector, we need not worry about the $z$ polarization, so we will confine our attention to the $x$ and $y$ polarizations in the detector plane.
Let $E_x^A$ represent the $x$ (dominant) component of the field that results from propagating the $A$ polarization through the system, and let $E_y^A$ represent the corresponding $y$ (cross) component.
Similarly, $E_x^B$ and  $E_y^B$, respectively, represent the cross and dominant fields that result from propagating the $B$ polarization through the system.
With these definitions, the time-dependent vector electric field at the detector is given by:
\begin{equation}
\bE(t) = \hat{\bx}[E_x^A a(t) + E_x^B b(t)] + \hat{\by}[E_y^B b(t) + E_y^A a(t)] \, ,
\label{eq: vector E(t)}
\end{equation}
where the harmonic term $\exp(j2\pi \nu t)$ has been dropped.

The intensity in the detector plane corresponding to Eq.~\eqref{eq: vector E(t)} is:
\begin{align}
I & = \overline{\bE(t) \cdot \bE^*(t)}   \label{eq: detector intensity 0} \\
  & = I_x^A + I_x^B + I_y^B + I_y^A \, , \label{eq: detector intensity} 
\end{align}
where $\cdot$ represents a scalar (i.e., dot) product and $ I_x^A =|E_x^A|^2$, $ I_x^B =|E_x^B|^2$, $ I_y^B =|E_y^B|^2$, $ I_y^A =|E_y^A|^2$.
The fact that the four terms in Eq.~\eqref{eq: vector E(t)} result in four terms in Eq.~\eqref{eq: detector intensity} with no cross terms is due to the fact that $\hat{\bx} \cdot \hat{\by}=0$ and the incoherence conditions in Eq.~\eqref{eq: incoherent envolopes}.   

\subsection{Including Two Linear Polarizers}\label{sec: polarizers}

The major hurdle to estimating the weak cross intensity, which could potentially be confounded with incoherent intensity, is that only under the dark hole condition is the cross intensity not necessarily overwhelmed by the dominant intensity.
Further, it is the authors' experience that it is not possible to significantly modulate the cross field for probing purposes (see below) without destroying the dark hole in the dominant intensity, which exists only under precise conditions.
However, two linear polarizers, one in the collimated entrance beam, and another just before the detector 
can change this dynamic substantially.
The concept is as follows:
\begin{itemize}
	\item The linear polarizer in the entrance beam only allows A-polarized light to pass (see Sec.\ref{sec: Intensity with CP}).
	This results in the dominant polarization being in the $x$-direction at the detector plane and the cross-polarization being in the $y-$direction [see Eq.~\eqref{eq: detector intensity}].
	
	\item The linear polarizer just before the detector only allows $y-$polarized light to pass, removing the dominant field; this allows the cross field to be measured.
\end{itemize}

The analysis provided below takes into account the fact that real polarizers have finite leakage, allowing small fractions, $\epsilon_1$ and $\epsilon_2$ (e.g., $10^{-5}$), of the attenuated intensity to pass. 
We model the Jones matrix of the linear polarizer placed in the entrance pupil, in the $AB$ coordinate system as:
\begin{equation}
H_{AB} = 
\left(\begin{array}{c c}
1 & 0 \\
0 & \sqrt{\epsilon_1}  
\end{array} \right) \, .
\label{eq: Jones AB}
\end{equation}
Similarly, we model the Jones matrix in the $xy$ coordinate system of the linear polarizer just before the detector as:
\begin{equation}
H_{xy} = 
\left(\begin{array}{c c}
\sqrt{\epsilon_2} & 0 \\
0 & 1  
\end{array} \right) \, .
\label{eq: Jones xy}
\end{equation}
More realistic Jones matrix values would be a bit less than unity (say, 0.95) for the transmission axis, but we ignore that to reduce clutter.
Applying these Jones matrices, Eq.~\eqref{eq: detector intensity} becomes:
\begin{equation}
I  = \epsilon_2 I_x^A + \epsilon_1 \epsilon_2 I_x^B + \epsilon_1 I_y^B + I_y^A \, . 
\label{eq: detector intensity w polarizers -1} 
\end{equation}
Thus, only the $I_y^A$ component of the cross intensity is not attenuated by the polarizers.
Importantly, under dark hole conditions, the dominant intensities, $I_x^A$ and $I_y^B$, are
severely attenuated as well.
As an example, one of the glass dichroic linear polarizers from Polarcor has an extinction coefficient $\delta < 10^{-5}$ and operates from about $\lambda = 960$ to $1160 \,$nm with less than $\lambda/4$ distortion.  
Of course, polarizers can be used in series to multiply their extinction factors, and this may be an attractive option.
The distortion from the polarizers is not included in this analysis, but in the Discussion section, we argue that they should be manageable without undue complication.

\subsection{Scalar Fourier Optics Approximations and Symmetry Considerations}\label{sec: Scalar and Symmetry}

Under the assumption that the light is unpolarized, the relationship between the scalar electric field $u$ and the  components of the vector field $(E_x, \, E_y, \, E_z)$ is simply 
\begin{equation}
E_z=0 \,, \: \:  \mathrm{and}\: E_x = E_y = u/\sqrt{2} \, ,
\label{eq: scalar optics}
\end{equation}
which is normalized so that the intensity value is consistent.   
Underlying standard Fourier optics principles, such as the Fourier transforming property of lenses, is the paraxial assumption that relies on the proximity of rays to the principal axis of the optical system.\cite{IntroFourierOptics}  
When the paraxial assumption does not hold, one must be prepared to reconsider Eqs.~\eqref{eq: scalar optics}.
Thus, if one is working in a regime accurately represented by Fourier optics, then, $E_x^A = E_y^B$ for the dominant fields, and $E_x^B = E_y^A = 0$ for the cross fields.
Since all model-based EFC formulations successfully use Fourier optics,\cite{Kasdin_EFC16b} it follows that $E_x^A$ and $E_y^B$ are essentially the same and that the dark hole corresponds to small values of both $|E_x^A|$ and $|E_y^B|$.
\emph{En revanche}, the cross fields $E_x^B$ and~$E_y^A$ arise due to non-paraxial effects, and simulations show that they are not substanially reduced in the process of creating the dark hole in the dominant field.

All of the calculations presented in this article are carried out without paraxial assumptions.
The simulation results in Figs.~\ref{fig: fields simple} and~\ref{fig: fields OAP} described above in Sec.~\ref{sec: Cross Pol Simulations} as well as those from the OAP-based coronagraph simulated in detail (described later) exhibited two symmetries for the dominant and cross fields.  Interestingly, even beams off-axis by angles  $100 \lambda/D$ also showed these same symmetries.  
The following equations express the two symmetries:
\begin{equation}
|E_x^A| = |E_y^B| \, ,  \: \: \mathrm{and} \; |E_x^B| = |E_y^A| \, ,
\label{eq: amplitude symmetry} 
\end{equation} 
the former of which is consistent with scalar field assumptions. 
Of course, Eqs.~\eqref{eq: amplitude symmetry} imply that $I_x^A = I_y^B$ and $I_x^B = I_y^A$.

Due to the above symmetry reasons leading to Eqs.\eqref{eq: amplitude symmetry}, the models in this article need only treat light linearly polarized in the $A$ direction entering the coronagraph
Thus, we will not calculate $E_x^B$ or $E_y^B$ and will drop the $A$ and $B$ superscripts.
We will take  $E_x$  and  $E_y$ to be the dominant and cross fields, respectively, in the detector plane.
With this simplification, Eq.~\eqref{eq: detector intensity w polarizers -1} becomes:
\begin{equation}
I  = (\epsilon_2 + \epsilon_1 )I_x + (\epsilon_1 \epsilon_2 + 1)I_y  \approx  (\epsilon_2 + \epsilon_1 )I_x + I_y \, . 
\label{eq: detector intensity w polarizers} 
\end{equation}
Below, we will leverage the extinction factor $(\epsilon_2 + \epsilon_1 )$ to allow us to apply DM commands that probe the cross field to the detriment of the dark hole in the dominant intensity.

\section{A Matrix-Based Coronagraph Model}\label{sec: models}

Let us represent the ($A-$polarized) field in the coronagraph DM plane, which is assumed to be located in the collimated beam from the entrance pupil, $\mathcal{A}$, as $U(\brho; \bc)$ where $\brho$ is the 2D spatial coordinate vector in the pupil plane and $\bc$ is a vector of DM actuator commands, known as the \emph{DM command} or, simply, \emph{command}.
Similarly, we represent the field in the detector plane as $\hat{\bx} E_x(\br; \bc) + \hat{\by} E_y(\br; \bc)$, where $E_x(\br; \bc)$ and $E_y(\br; \bc)$ are dominant and cross fields, respectively. 
Coronagraphs are linear optical systems in the sense that the beam does not pass through nonlinear gain media such as laser cavities.
Due to linearity, we have the relationships for the dominant and cross fields:
\begin{align}
E_x(\br; \bc) & =  \int_\mathcal{A} \rd \brho \, \mathcal{K}_x(\br,\brho)U(\brho; \bc) \, , \: \: \: \mathrm{and}
\label{eq: kernel X} \\
E_y(\br; \bc) & = \int_\mathcal{A} \rd \brho \, \mathcal{K}_y(\br,\brho)U(\brho; \bc) \, ,
\label{eq: kernel Y}
\end{align}
\begin{wrapfigure}{l}{0.45\textwidth}
	\includegraphics[width=0.45\textwidth]{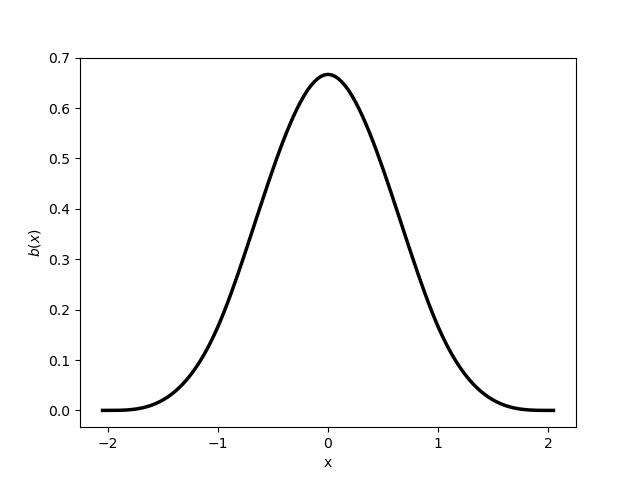}
	\caption{\small A cubic B-spline (CBS) basis function in 1D.
		Here, the width parameter $\delta  = 1$.  This function is supported on the interval $(-2\delta, \, 2\delta)$.\cite{UnserSpline}}\label{fig: CBS1D}
\end{wrapfigure}
where $\mathcal{K}_x(\br,\brho)$ and $\mathcal{K}_y(\br,\brho)$ are complex-valued kernel functions corresponding to the coronagraph optical system.
Eqs.~\eqref{eq: kernel X} and~\eqref{eq: kernel Y} should be taken as true statements with no approximations with the understanding that the detector fields $E_x(\br; \bc)$ and~$E_y(\br; \bc)$, the kernels $\mathcal{K}_x(\br,\brho)$ and~$\mathcal{K}_y(\br,\brho)$ and the ($A$-polarized) input field $U(\brho; \bc)$ are never exactly known.
The true fields present at detector pixel $m$ are $E_x(\br_m; \bc)$ and $E_y(\br_m; \bc)$.
When we consider all $M$ detector pixels at once, the corresponding vectors, each with $M$ components, are $\bff_x(\bc)$ and $\bff_y(\bc)$.

Matrix-based numerical models of the coronagraph optical system can take advantage of the linearity in Eqs.~\eqref{eq: kernel X} and~\eqref{eq: kernel Y}.
For example, in their study of millisecond exoplanet imaging from ground-based platforms with adaptive optics, Rodack and Frazin [\citenum{RodackFrazin_JOSAA21}, \citenum{FrazinRodack_JOSAA21}] divided both the entrance pupil and the detector plane into pixels and modeled the system with a matrix equivalent of the kernel function $\mathcal{K}_x(\br,\brho)$.
In the single DM coronagraph model presented here (described in more detail later), the field in the entrance pupil of the system is a plane wave modified by the DM.
We simulate a DM with $\sqrt{N} \times \sqrt{N}$ actuators, each of which has an influence function modeled by a 2D cubic B-spline basis (CBS) function.\cite{UnserSpline} 
As depicted in 1D in Fig.~\ref{fig: CBS1D}, the CBS looks much like a Gaussian.
The CBS has a width parameter $\delta$ that is analogous to the standard deviation of a Gaussian.
Unlike Gaussians, CBSs have finite extent in the spatial domain, yet they are well-behaved with usually negligible ringing under Fourier transformation.
The 2D CBS is simply the tensor product of the 1D CBS with itself, i.e., $\eta(\brho) = b(x)b(y)$, where $\brho = (x, \, y) $ and $b(x)$ is the 1D CBS function. 
We model the field corresponding to the light reflecting off of the DM at the location $\brho$ as:
\begin{equation}
u(\brho; \, \bc) = \sum_{n=0}^{N-1} \eta(\brho - \brho_n) \exp (j c_n) 
\label{eq: DM height}
\end{equation}
where $\brho_n$ is the location of the center of the $n$\underline{th} DM actuator,
and the command vector $\bc$, of length $N$, contains the phases $c_0 , \, \dots, \, c_{N-1}$ applied to the DM influence functions $\eta(\brho - \brho_0) , \, \dots, \, \eta(\brho - \brho_{N-1})$.
For convenience, although we call $c_n$ the command for actuator $n$ in this article, in reality, it is the phase corresponding to the command.
Thus, assuming normal incidence on the DM, $c_n = 4 \pi (h_n/\lambda)$, in which $h_n$ is the commanded height for actuator $n$ and $\lambda$ is the wavelength.
This somewhat nonstandard nomenclature will serve to reduce clutter.

In these simulations, the field in the entrance pupil was taken to be a 2D cubic B-spline basis function centered on a DM actuator, and the Virtual Lab Fusion (VLF) simulation package from LightTrans, Inc. performed non-paraxial propagation through the coronagraph to determine the resulting dominant and cross fields in all $M$ detector pixels.
Doing this for all $N$ DM actuators results in the $M \times N$ complex-valued matrices $\tilde{\bD}_x$ and $\tilde{\bD}_y$. 
We denote the numerical model of the dominant and cross fields in the detector plane as $\bv_x(\bc)$ and $\bv_y(\bc)$, each of which has $M$ components.
Then, the model is captured by the matrix-vector equations:
\begin{align}
\bv_x(\bc) & = \tilde{\bD}_x \bu(\bc) \: \: \: \mathrm{and} \label{eq: MVM X -1} \\
\bv_y(\bc) & = \tilde{\bD}_y \bu(\bc) \, ,  \label{eq: MVM Y -1}
\end{align}
where $\bu(\bc)$ , is a vector of $N$ phasors with the $n$\underline{th} component being $\exp(j c_n)$, so that $\bu(\bc_1 + \bc_2) = \bu(\bc_1) \circ \bu(\bc_2)$, where $\circ$ is the Hadamard (i.e. element-wise) product.
Eqs.~\eqref{eq: MVM X -1} and~\eqref{eq: MVM Y -1} are the model analogs of Eqs.~\eqref{eq: kernel X} and~\eqref{eq: kernel Y}.
While VLF took a couple of days to create the matrices $\tilde{\bD}_x$ and $\tilde{\bD}_y$, the matrix-vector multiplications shown in Eqs.~\eqref{eq: MVM X -1} and~\eqref{eq: MVM Y -1} are fast and ideally suited to iterative algorithms.\cite{MatrixProp_Shakir15, Frazin_JOSAA2018}

\subsection{The Dark Hole}\label{sec: Dark Hole}

We start under the assumption that an EFC loop has already succeeded in creating a dark hole in a region on the detector region $\mathcal{D}$, which is a set of pixels, and let $\bc_0$ represent the command corresponding to the dark hole configuration.
In this article, we will only consider DM commands in the vicinity $\bc_0$, so it makes sense to economize the notation by redefining $\bc$ to be the difference between the DM command and $\bc_0$.
Replacing $\bc$ with $\bc_0 + \bc$ in Eqs.~\eqref{eq: MVM X -1} and~\eqref{eq: MVM Y -1} achieves this change of origin.
Thus, we must replace $\bu(\bc)$ with $\bu(\bc_0 + \bc)   = \bu(\bc_0) \circ \bu(\bc) $.
Note that the matrix-vector product $\tilde{\bD}_x [\bu(\bc_0) \circ \bu(\bc)] = [\tilde{\bD}_x \, \mathrm{diag} \big(\bu(\bc_0)\big) ]\bu(\bc)$, where $\mathrm{diag}\big(\bu(\bc_0)\big)$ is a $N \times N$ diagonal matrix with the $n$\underline{th} entry $\exp(j c_{0n})$.
Next, we define $\bD_x \equiv \tilde{\bD}_x \, \mathrm{diag} \big(\bu(\bc_0)\big)$ so that Eqs.~\eqref{eq: MVM X -1} and~\eqref{eq: MVM Y -1} become:
\begin{align}
\bv_x(\bc) & = \bD_x \bu(\bc) \: \: \: \mathrm{and} \label{eq: MVM X} \\
\bv_y(\bc) & = \bD_y \bu(\bc) \, . \label{eq: MVM Y}
\end{align}
Eqs.~\eqref{eq: MVM X} and~\eqref{eq: MVM Y}  contain no approximations not already present in Eqs.~\eqref{eq: MVM X -1} and~\eqref{eq: MVM Y -1}.
Just to be clear, $\bv_x(\boldsymbol{0}) = \bD_x \b1 $ and $\bv_y(\boldsymbol{0}) =  \bD_y \b1  $, in which $\boldsymbol{0}$ and~$\b1$ are vectors of all zeros and ones, are the dominant and cross fields attained in the dark hole configuration.

Without confining ourselves to linearized analysis, we can replace  $\bu(\bc)$ with  $ \b1 + \big(\bu(\bc) -\b1 \big)$ in Eqs.~\eqref{eq: MVM X} and~\eqref{eq: MVM Y} and arrive at :
\begin{align}
\bv_x(\bc) & = \bv_x(\boldsymbol{0}) + \bD_x \big(\bu(\bc) - \b1 \big)\: \: \: \mathrm{and} \label{eq: MVM X 2} \\
\bv_y(\bc) & = \bv_y(\boldsymbol{0}) + \bD_y \big(\bu(\bc) - \b1 \big) \, , \label{eq: MVM Y 2}
\end{align}
where $ \bv_x(\boldsymbol{0}) = \bD_x \b1$ and~$ \bv_y(\boldsymbol{0}) = \bD_y \b1$ are the model fields at the dark hole configuration.
Similarly, the true field can be expressed as:
\begin{align}
\bff_x(\bc) & = \bff_x(\boldsymbol{0}) + \frac{\partial \bff_x}{\partial \bu} \big(\bu_\rt(\bc) - \b1 \big)\: \: \: \mathrm{and} \label{eq: true separation X} \\
\bff_y(\bc) & = \bff_y(\boldsymbol{0}) + \frac{\partial \bff_y}{\partial \bu} \big(\bu_\rt(\bc) - \b1 \big) \, , \label{eq: true separation Y}
\end{align}
in which $\bu_\rt(\bc)$ is the unknown true field reflecting off of the DM, and the unknown Jacobians, ${\partial \bff_x}/{\partial \bu}$ and~${\partial \bff_y}{\partial \bu}$, are constants due to the linearity of Eqs.~\eqref{eq: kernel X} and~\eqref{eq: kernel Y}.
At this point in these developments, we have an exactly known numerical model, as summarized in Eqs.~\eqref{eq: MVM X 2} and~\eqref{eq: MVM Y 2}.
In contrast, Eqs.~\eqref{eq: true separation X} and~\eqref{eq: true separation Y} consist of unknown quantities, but their virtue is accurately representing the true optical system.
We can hybridize Eqs.~\eqref{eq: MVM X 2} and~\eqref{eq: MVM Y 2} and Eqs.~\eqref{eq: true separation X} and~\eqref{eq: true separation Y} to arrive at what henceforth will be referred to as the \emph{hybrid equations}:
\begin{align}
\bff_x(\bc) & \approx \bff_x(\boldsymbol{0}) +  \bD_x \big(\bu(\bc) - \b1 \big) \: \: \: \mathrm{and} \label{eq: hybrid X} \\
\bff_y(\bc) & \approx \bff_y(\boldsymbol{0}) +  \bD_y \big(\bu(\bc) - \b1 \big)\, , \label{eq: hybrid Y}
\end{align}
which are nonlinear because $\bu(\bc)$ is nonlinear.
The validity of the hybrid equations relies on the hope that replacing $\big(\partial \bff_x / \partial \bu \big) \big(\bu_\rt(\bc) - \b1 \big)$ with  $\bD_x \big(\bu(\bc) - \b1 \big)$ for the dominant field, and similarly for the cross field, are only misdemeanor offenses, at least when $\bc$ is sufficiently small.
The fields, $\bD_x \big(\bu(\bc) - \b1 \big)$ and  $\bD_y \big(\bu(\bc) - \b1 \big)$ and their linearized versions (see below), are known as \emph{probe fields} or just \emph{probes}.

Recalling that the $n$\underline{th} element of $\bu(\bc)$ is $\exp( j c_n)$, it follows that the linearization of $\bu(\bc)-\b1$ is simply $j\bc$. 
Working from  Eqs.~\eqref{eq: hybrid X} and~\eqref{eq: hybrid Y}, the \emph{linearized hybrid equations} are:
\begin{align}
\bv_x(\bc) & \approx \bv_x(\boldsymbol{0}) + j\bD_x \bc \: \: \: \mathrm{and} \label{eq: MVM X linearized} \\
\bv_y(\bc) & \approx \bv_y(\boldsymbol{0}) + j\bD_y \bc \, . \label{eq: MVM Y linearized}
\end{align}
Note that all EFC methods, except for the model-free one of Haffert et al. and the phase-retrieval-type approach of Malbet, Yu and Shao,\cite{Haffert-ModelFree_AA23, Malbet_EFC95} rely on the validity of the linearized hybrid equation in the dominant field.\cite{Kasdin_EFC16b, Traub_Nulling06, Potier_BordeTraub_SPIE2024}  
It is likely that the accuracy of the linearized hybrid equation improves as the iterations approach the dark hole state.

\section{Estimation of the Cross-Polarization Electric Field}\label{sec: Probing}

\begin{wrapfigure}{r}{0.45\textwidth}
\vspace{-3mm}
	\includegraphics[width=0.45\textwidth]{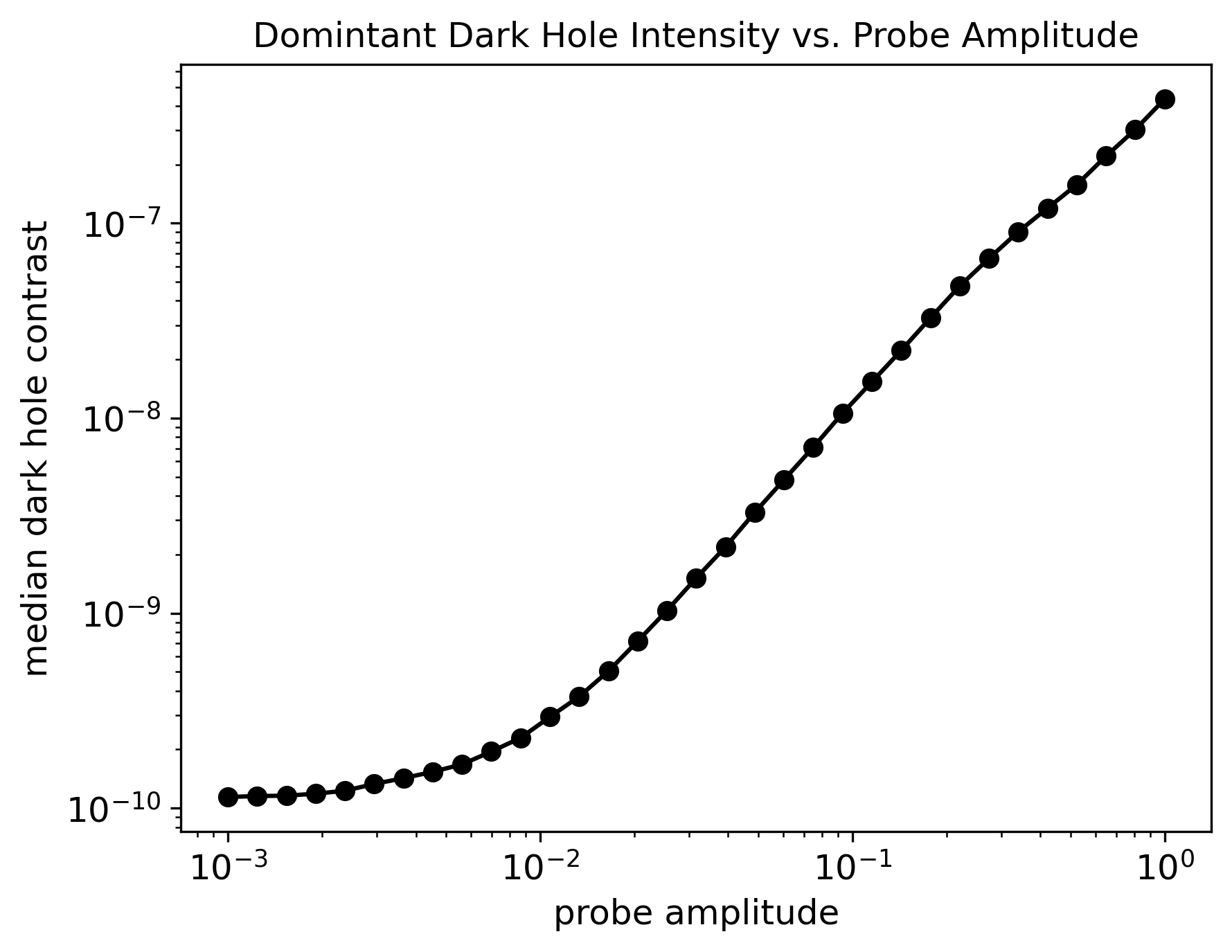}
	\caption{\small Dark Hole (dominant) intensity vs. probe amplitude for one of the probes used to estimate the cross field.  A probe amplitude of unity corresponds to the amplitude used in the intensity measurement for estimating the cross field.  This does not include the linear polarizers in the beam path.  Including the polarizers amounts to multiplying the $y-$axis values by $(\epsilon_1+\epsilon_2)$, as per Eq.~\eqref{eq: detector intensity w polarizers}.  See Sec.~\ref{sec: Probing} for details.}
	\label{fig: IntVsAmp}
	\vspace{0mm}
\end{wrapfigure}

The primary difficulty with probing the cross field is that the DM commands that probe the cross field effectively compromise the dark hole.
Under the assumption of linearity, it is possible to satisfy the dominant dark hole condition at a subset of pixels while probing the cross field, but, in practice, the linearity assumption is not satisfied for effective cross probes.
Fig.~\ref{fig: IntVsAmp}, which shows $I_x$ as a function of the amplitude of a probe command, depicts this concept.
One of the probes chosen for the simulated estimations of the cross field (described later) was used to create Fig.~\ref{fig: IntVsAmp}.
A probe amplitude of unity corresponds to the probe that was used for a measurement as part of the estimation process.
The figure, which does not include the linear polarizers, shows that the dark hole is progressively destroyed as the probe amplitude increases.
The linear polarizers can be taken into account by multiplying the $y-$axis by $(\epsilon_1 + \epsilon_2)$, as per Eq.~\eqref{eq: detector intensity w polarizers}.

Starting with the hybrid Eqs.~\eqref{eq: hybrid X} and~\eqref{eq: hybrid Y}, let the dominant and cross field probes be represented by the vectors $\bp_x(\bc)$ and $\bp_y(\bc)$, which can represent either the linearized (i.e., $j\bD_x \bc $ for the dominant field) or nonlinear variant [i.e., $\bD_x \big(\bu(\bc) - \b1 \big)$ for the dominant field].
Using $\bp_x(\bc)$ for the dominant probe field, multiplying the hybrid equation for the dominant field by its complex conjugate yields:
\begin{align}
\bi_x(\bc) = \: \: & \bh +  |\bff_x(\boldsymbol{0})|^2  + |\bp_x(\bc) |^2   \nonumber \\
& - \Im \big( \bp_x(\bc)  \big) \circ \Re(\bff_x(\boldsymbol{0}))  +
\Re \big( \bp_x(\bc)  \big) \circ \Im(\bff_x(\boldsymbol{0}) \big) \, ,
\label{eq: intensity probe X}
\end{align}
where $\Re$ and~$\Im$ indicate the real and imaginary parts, $\bh$ is the incoherent component, which is assumed independent of the DM command,\footnote{This is a good approximation so long as the DM modulations do not significantly alter the Strehl ratio.}
 and$\, \circ \,$ represents the Hadamard (i.e., element-wise) product.
For the cross field, there is an equation similar to Eq.~\eqref{eq: intensity probe X}:
\begin{align}
\bi_y(\bc) = \: \: &  |\bff_y(\boldsymbol{0})|^2  + |\bp_y(\bc) |^2   \nonumber \\
& - \Im \big( \bp_y(\bc)  \big) \circ \Re(\bff_y(\boldsymbol{0}))  +
\Re \big( \bp_y(\bc)  \big) \circ \Im(\bff_y(\boldsymbol{0}) \big) \, ,
\label{eq: intensity probe Y}
 \end{align}
Note that Eq.~\eqref{eq: intensity probe Y} has no incoherent component, $\bh$.
This is because the incoherent source is assumed to be weak, making its cross-polarization signal negligible.

\subsection{Regression Equations}\label{sec: Regression}

Simply put, our aim is to use the hybrid equations [i.e., Eqs.~\eqref{eq: hybrid X} and~\eqref{eq: hybrid Y}] to model probed intensities and estimate the real and imaginary parts of the unprobed cross field, $\bff_y(\boldsymbol{0})$.
Note that the use of the linear polarizers will extinguish the incoherent component $\bh$ in Eq.~\eqref{eq: intensity probe X}, the estimation of which could be done subsequently to obtaining an estimate of the cross polarization.
The usual pairwise probing procedure in EFC methods uses DM commands that are small enough to allow estimation of the image plane field via linear regression.
However, the DM commands we select to probe the cross field are too large to safely apply linear regression and we apply a nonlinear regression procedure described below.
Nonlinear regression for the dark hole problem dates back to Malbet, Yu and Shao,\cite{Malbet_EFC95} 
but they solved for a pupil-plane phase error, not the field in the focal plane, which makes their approach more closely related to classical phase retrieval.\cite{PhaseRetrieval_review2015}

Measuring the intensity with $P$ different probe commands, $\bc_1 , \, \dots, \, \bc_P$ enables the regression on the real and imaginary parts of the cross field in the dark hole, $\bff_y(\boldsymbol{0})$.
The corresponding dominant field $\bff_x(\boldsymbol{0})$ could be estimated as well (with a sufficient number of probes), but the linear polarizers make this difficult, so it is better to estimate it without the polarizers in place.  
In the Discussion section,  we outline a practical approach to adjust the dark hole command to account for the wavefront error introduced by the polarizers.
For simplicity, we consider only estimating the real and imaginary parts of the cross field, $\bff_y(\boldsymbol{0})$.
When we apply the linear polarizers as described in Sec.~\ref{sec: polarizers}, the measured intensity in the detector pixels with probe command $\bc_i$ is:
\begin{equation}
\by_i = \epsilon \bi_x(\bc_i) + \bi_y(\bc_i) + \bn_i \, ,  \; 1 \leq i \leq P \, ,
\label{eq: y_i simple} 
\end{equation}
where $\epsilon = \epsilon_1 + \epsilon_2$, and $\bn_i$ is a vector of length $M$ corresponding to noise that arises from the usual culprits, e.g., photon counting statistics, readout noise and thermal (dark) noise.
Using Eqs.~\eqref{eq: intensity probe X} and ~\eqref{eq: intensity probe Y} in Eq.~\eqref{eq: y_i simple}, we get
\begin{align}
\by_i  = \epsilon \big[  & |\bff_x(\boldsymbol{0})|^2  + |\bp_x(\bc_i) |^2   - \Im \big( \bp_x(\bc_i)  \big) \circ \Re(\bff_x(\boldsymbol{0}))  +
\Re \big( \bp_x(\bc_i)  \big) \circ \Im(\bff_x(\boldsymbol{0}) \big)    \big] \nonumber \\
+ \: &  |\bff_y(\boldsymbol{0})|^2  + |\bp_y(\bc_i) |^2  - \Im \big( \bp_y(\bc_i)  \big) \circ \Re(\bff_y(\boldsymbol{0}))  +\Re \big( \bp_y(\bc_i)  \big) \circ \Im(\bff_y(\boldsymbol{0}) \big) + \bn_i \, , 
\label{eq: y_i full}
\end{align}
where we have dropped $\epsilon \bh$ since this is a negligible quantity, and $ 1 \leq i \leq P$.
Since we are taking $\bff_x(\boldsymbol{0})$ to be known (or at least accurately estimated already), there are $P$ measurement equations of the form shown in Eq.~\eqref{eq: y_i full}, from which we can estimate two unknowns (per pixel),  $\Re\big(\bff_y(\boldsymbol{0})\big)$ and $ \Im\big(\bff_y(\boldsymbol{0})\big)$.
In these simulations, which use count rates one might expect from a 100 mW supercontinuum laser, each measurement had hundreds or thousands of counts per pixel, which mean we can neglect readout noise, and we can take the measured intensity $\by_i$ as Poisson distributed to account for photon counting statistics.
Because the measurements are statistically independent from one pixel to the next, the analysis to follow need only consider a single pixel in the dark hole, henceforth ``the'' pixel.
Let $k_i$ represent the number of counts measured with probe $i$ in the pixel, so $k_i$ is one of the components of $\by_i$.
The two-component vector $\balpha = (\alpha_1, \, \alpha_2)$ will represent the values obtained by $\Re(\bff_y(\boldsymbol{0})$ and $ \Im(\bff_y(\boldsymbol{0})$ in the pixel, which are the quantities we will estimate.
Let $s_i(\balpha)$ be the (unknown because $\balpha$ is unknown) true intensity associated with probe $i$, for which we have the measured counts $k_i$.
Thus, we have:
\begin{equation}
s_i(\balpha) = \epsilon \, i_x(\bc_i) + 
 \alpha_1^2 + \alpha_2^2  + |p_y(\bc_i) |^2  - \Im \big( p_y(\bc_i)  \big) \alpha_1  +\Re \big( p_y(\bc_i)  \big) \alpha_2 \, ,
 \label{eq: s_i}
\end{equation}
where $ \alpha_1^2 + \alpha_2^2 = { \small |\bff_y(\boldsymbol{0})|^2} $, $i_x(c_i)$ is the value $\bi_x(c_i)$ obtains in the pixel [see Eq.~\eqref{eq: intensity probe X}], and $p_y(\bc_i)$ is the value obtained by the cross probe field $\bp_y(\bc)$ in the pixel.
Assuming $s_i(\balpha)$ is in photon units, the probability of obtaining $k_i$ counts under the Poisson distribution is  
$[s_i(\balpha)]^{k_i}\exp[-s_i(\balpha)] /k_i$! 
The negative log-likelihood is:
\begin{equation}
\mathcal{Q}(\balpha) = \sum_{i=0}^{P}\left[ s_i(\balpha) - k_i \ln\big( s_i(\balpha) \big) \right]      
\label{eq: Poisson LL}
\end{equation}
where $i=0$ corresponds to an unprobed measurement (i.e., the dark hole state), and the $\ln (k_i!)$ term has been dropped since it does not depend on the estimated quantities, $\balpha$.
The estimate of $\balpha$ corresponds to a minimum of $\mathcal{Q}(\balpha)$, hopefully a global minimum.
Using the symbol $\nabla$ to represent the partial derivatives with respect to the components of $\balpha$, it is useful to calculate $\nabla \mathcal{Q}(\balpha)$ for both employing gradient-based minimization algorithms and for calculating the Fisher information matrix (FIM) and its inverse: the Cram\'er-Rao bound.\cite{van_trees1968}
Then,
\begin{equation}
\nabla \mathcal{Q}(\balpha) = \sum_{i=0}^{P} \left\{ \nabla s_i(\balpha) \left[1 - \frac{k_i}{s_i(\balpha)} \right] \right\} \, ,
\label{eq: grad Poisson LL}
\end{equation}
where $ \nabla s_i(\balpha) $ can be found from Eq.~\eqref{eq: s_i}.
The Hessian, $ \nabla \nabla \mathcal{Q}(\balpha) $, which is useful in Newton minimization methods, is easily obtained as well. 
The Cram\'er-Rao bound is a lower bound on the estimate error covariance for an unbiased estimator, and it proved to be an accurate estimator of the errors in the simulations shown below.
A few lines of algebra show that under the Poisson distribution, the Fisher information matrix (FIM), denoted by $\bF$, takes an extraordinarily simple form:
\begin{equation}
\bF = \sum_{i=0}^{P}   \frac{1}{s_i(\balpha)} \nabla s_i(\balpha)^\rT  \nabla s_i(\balpha) \, ,
\label{eq: FIM}
\end{equation}
where $^\rT$ denotes the transpose operator and $\nabla s_i(\balpha)$ is taken to be a column vector, thus, Eq.~\eqref{eq: FIM} implies the outer product of the gradient vectors.

\subsection{Nonlinear Estimation}\label{sec: estimation}

The estimate of $\balpha$, denoted as $\hat{\balpha}$, is taken to be the least of the multiple local minima of $\mathcal{Q}(\balpha)$ found by applying local minimization algorithms at multiple starting points.  
The unprobed, i.e., $i=0, \, c_i = \boldsymbol{0}$, measurement is critical for choosing the set of starting points for the algorithms.
Referring to Eq.~\eqref{eq: s_i}, first, notice that $p_y(\boldsymbol{0})=0$ [see the remarks just before Eq.~\eqref{eq: intensity probe X}]. 
Next, recall that the unprobed measurement corresponds to the dark hole state of the dominant field, which may have an intensity of, say, $10^{-10}$ in contrast units, without taking the linear polarizers into account.  
Including linear polarizers with an extinction factor of, for example, $\epsilon = 10^{-5}$, then the first term in Eq.~\eqref{eq: s_i} is $\epsilon i_x(\boldsymbol{0})$ is, perhaps, $10^{-15}$.
Therefore, the unprobed intensity is almost entirely the cross intensity, so, $s_0 \approx \alpha_1^2 + \alpha_2^2$, which may have a value somewhere around $10^{-13} - 10^{-11}$.    
The corresponding measured number of counts is $y_0$, so we have $y_0 \approx  \alpha_1^2 + \alpha_2^2$.
In the simulations below, we performed $N_\mathrm{opt}$ local minimizations with the $l$\underline{th} starting point being:
\begin{equation}
(\alpha_{1l}, \alpha_{2l}) = \zeta \sqrt{y_0} \big(\cos(\phi_l), \, \sin(\phi_l)  \big) \, , 
\theta_l = \frac{2  \pi l}{ N_\mathrm{opt}}    \, , 0 \leq l <  N_\mathrm{opt}  \, , 
\label{eq: opt starting points}
\end{equation} 
where $\zeta$ accounts for the scaling between $\sqrt{\mathrm{contrast}}$ units and $\sqrt{\mathrm{photon \, count}}$ units of the field.
In the simulations below,  $N_\mathrm{opt} = 24$ was used but a smaller number would likely have sufficed, and each of the $N_\mathrm{opt}$ starting points, we applied the conjugate gradient algorithm, with the analytical gradient from Eq.~\eqref{eq: grad Poisson LL}.
The estimate error covariance matrix was taken to be the Cram\'er-Rao bound matrix, which is the inverse of the FIM shown in Eq.~\eqref{eq: FIM}.
The so-obtained error estimates are consistent with differences between the estimates and true values in these simulations.
These nonlinear estimates and the error bar calculations are carried out rather quickly on a pixel-by-pixel basis and do not represent a significant computational cost.

\begin{figure}[h]
	\includegraphics[width=0.99\textwidth]{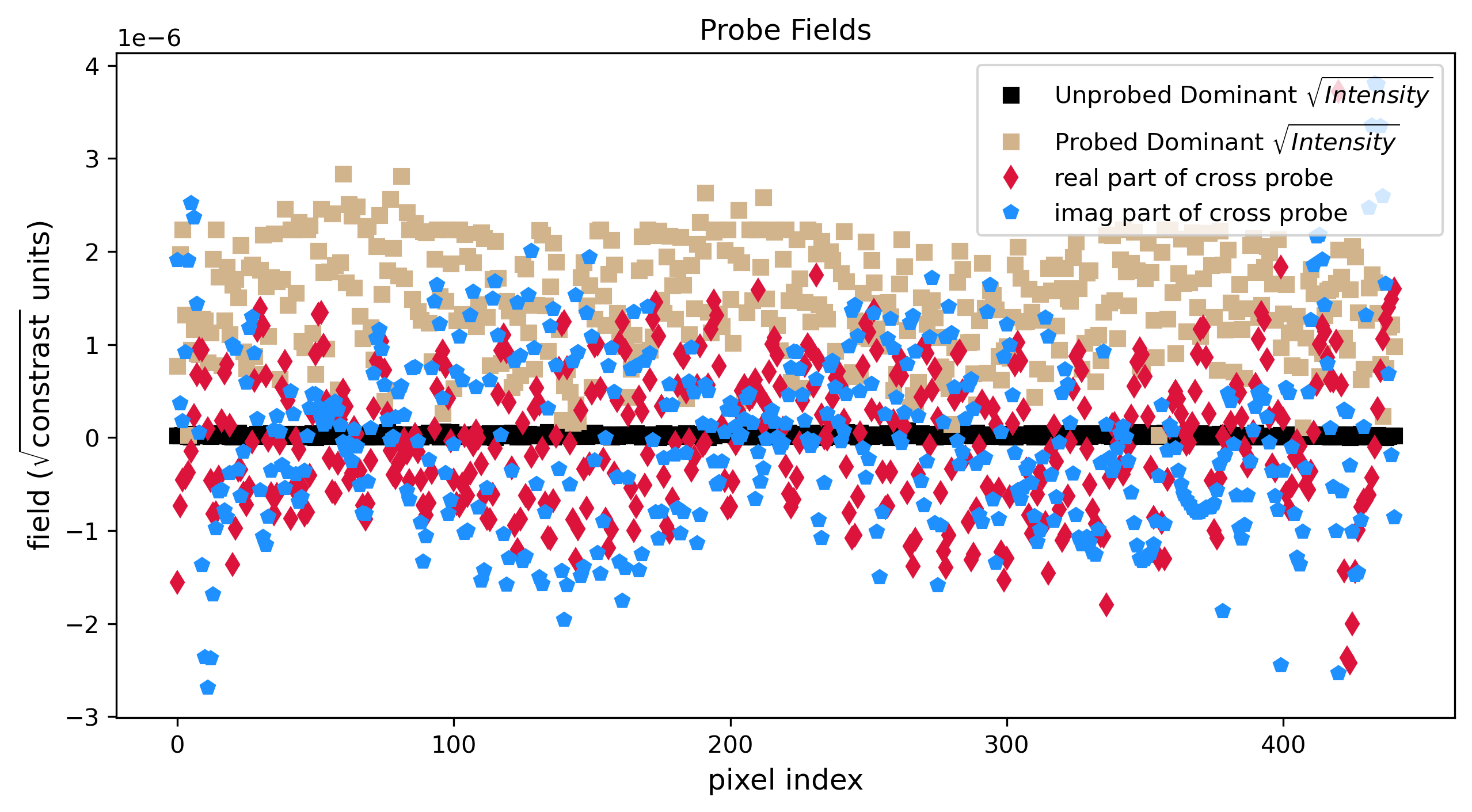}
	\caption{\small A simulation example of the dominant intensity and cross fields associated with one of the probes used for estimating the cross field in the dark hole.
		The $x-$axis is the pixel index within the dark hole (which has 441 pixels), and the $y-$axis is the electric field in units of $\sqrt{\mathrm{contrast}}$.
		The red diamonds and blue pentagons correspond to the real and imaginary probe fields, respectively. 
		The black and beige squares correspond to the square-root of the dominant intensity.  The black squares are the values obtained with no probe applied (i.e., the dark hole state), and the beige squares are obtained when this probe is applied.   The dominant $\sqrt{\mathrm{intensity}}$ values include a linear polarizer extinction factor of $\sqrt{\epsilon_1 + \epsilon_2}\, = \, 2.24 \times 10^{-3}$.  }
	\label{fig: ManyFields}
\end{figure}

\subsection{Finding Informative Probes}\label{sec: Finding Probes}

The primary difficultly in probing the cross field is that the effective cross probe commands destroy the dark hole in the dominant field, as shown in Fig.~\ref{fig: IntVsAmp} and Fig.~\ref{fig: ManyFields}, which leads to the cross intensity being swamped by the dominant intensity.
This is unavoidable; hence, the use of the linear polarizers to remove most of the dominant intensity.
The goal in this section is to find three probe commands that, when combined with the unprobed measurement, allow accurate estimation of the cross field in each pixel of the dark hole.
The strategy employed in these simulations consists of several parts:
\begin{enumerate}
	\item Find a set of $L > 3$  DM commands that correspond to local minima of a non-quadratic cost function that enforces desirable properties.  (The simulations below used $L=61$.)  The cost function is given below in Sec.~\ref{sec: Probe Cost Fcn}.
	\item Given the $L$ candidate solutions, evaluate the Cram\'er-Rao bound for \linebreak all $\gamma = L!/(L-3)!3!$ possible solution triples.
	Note that the Cram\'er-Rao bound, $\bC$, is a $2\times 2$ matrix, evaluated for each dark hole pixel.
	\item At each dark hole pixel, determine $\nu = \,$max[diag($\bC$)], where the diag operator extracts the diagonal elements of a matrix, to get the worse of the error bounds on the estimates of the real and imaginary parts.
	\item For each of the $\gamma$ solution triples, find the ``worst,'' i.e., the largest, value of $\nu$ among all the dark hole pixels.  Let us denote this value as $\nu_\mathrm{max}$.  $\nu_\mathrm{max}$ is the worst error bound of any of the estimated quantities in the dark hole.
	\item Of the $\gamma$ solution triples (see step 2 above), pick the ``best of the worst'', i.e., the solution triple that has the smallest value of $\nu_\mathrm{max}$.  This is the solution triple chosen for the measurements.
\end{enumerate}
Fig.~\ref{fig: ManyFields} shows the result of one of the probe commands chosen using the simulations described in more detail later.
The $x-$axis indexes the 441 pixels within the dark hole.
The values in the figure include the effect of the polarizers on the dominant intensity.
The dominant $\sqrt{\mathrm{intensity}}$ in the dark hole state, i.e., the unprobed state, is depicted with black squares, which crowd the $x-$axis in the figure.
The beige squares represent the dominant $\sqrt{\mathrm{intensity}}$ when the probe is applied, and it can be seen that in most pixels this  $\sqrt{\mathrm{intensity}}$ is within a factor of several of the real and imaginary parts of the probe fields.  
Clearly, without the linear polarizers in place, which multiply the dominant $\sqrt{\mathrm{intensity}}$ by about $0.0023$, the cross probe fields would be completely overwhelmed by the dominant intensity.
Of course, similar plots could have been provided for the other two probes chosen in this process.
At a qualitative level, they do not look much different, but Cram\'er-Rao analysis in the above procedure ensures that the real and imaginary parts of the probe fields are sufficiently different to enable accurate estimates.

\subsubsection{The Probe Cost Function}\label{sec: Probe Cost Fcn}
The $L$ candidate solutions mentioned in step 1 are local minima of a cost function defined in this section.
The cost function, $\mathcal{C}(\bc)$, includes several terms to encourage desirable properties in the probe command:
\begin{align}
\mathcal{C}(\bc) & = - \, \mathcal{C}_\mathrm{ps}(\bc) +  \mathcal{C}_\mathrm{dom}(\bc) +  \mathcal{C}_\mathrm{amp}(\bc) , \,
\label{eq: probe cost total} \, , \; \mathrm{where} \\
\mathcal{C}_\mathrm{ps}(\bc) &  =  \sum_\mathrm{pixels} |p_y(\bc)|^2  
\label{eq: cost probe strength} \\
\mathcal{C}_\mathrm{dom}(\bc) & =  \beta_\mathrm{dom} \sum_\mathrm{pixels} \big(i_x(\bc) -
    t_\mathrm{dom}\big) T \big(i_x(\bc), \, t_\mathrm{dom}\big) 
\label{eq: cost dom intensity} \\
\mathcal{C}_\mathrm{amp}(\bc) & =\beta_\mathrm{amp}\sum_{m=0}^{M-1}  \left\{   \big(c_m - t_\mathrm{amp} \big) T\big(c_m, \, t_\mathrm{amp}\big)
                  +                          \big(-c_m - t_\mathrm{amp} \big) T\big(-c_m, \, t_\mathrm{amp}\big) \right\}
\label{eq: cost amplitude}
\end{align}
where the sum over pixels refers to those in the dark hole, $T(\mu,\nu)$ is an indicator function that is zero when $\mu \leq \nu$ and unity otherwise, and $c_m$ is the DM command corresponding to actuator $m$.  
The coefficients $\beta_\mathrm{dom}$ and $\beta_\mathrm{amp}$, as well as the threshold values $t_\mathrm{dom}$ and $t_\mathrm{amp}$, are tuning parameters chosen via a trial-and-error procedure.
In Eq.~\eqref{eq: probe cost total},  $\mathcal{C}_\mathrm{ps}(\bc)$ rewards the strength of the cross probe intensity $ |\bp_y(\bc)|^2 = |\bD_y \big(\bu(\bc) - \mathbf{1} \big)|^2$ in its full nonlinear form.  
The function $\mathcal{C}_S\mathrm{dom}(\bc)$ penalizes the dominant intensity, $i_x(\bc)$, when it exceeds the threshold value $t\mathrm{dom}$; it does not account for the effect of the linear polarizers.
Note that $i_x(\bc)$ is not linearized here.
The function $\mathcal{C}_\mathrm{amp}(\bc)$ penalizes the DM actuator commands that exceed the threshold value  $t_\mathrm{amp}$, either positively or negatively.  The value used here was $t_\mathrm{amp} = \pi/4$.
This serves as a form of regularization.

Eq.~\eqref{eq: probe cost total} is non-quadratic in its argument $\bc$, and with $33^2=1089$ actuators in these simulations, it is not low-dimensional, so, it is not surprising that it has many local minima.  
However, for our purposes, having many local minima is a useful feature because each one is a possible probe choice, allowing us to select the most effective ones via the procedure outlined at the beginning of Sec.~\ref{sec: Finding Probes}.
The matrix-based representation of the optical system in Eqs.~\eqref{eq: MVM X} and~\eqref{eq: MVM Y} enables rapid evaluation of $p_y(\bc)$ and $i_x(\bc)$ in Eqs.~\eqref{eq: cost probe strength} and~\ref{eq: cost dom intensity}.
This rapid evaluation capability is crucial because, even with the supplied analytical gradient, the conjugate gradient algorithm required approximately 500,000 function evaluations to converge to each local minimum.
Each local minimum resulted from conjugate gradient steps starting with random values for the 1089 actuator commands.

\begin{figure}[t]
	\centering
	\includegraphics[width=1.1\textwidth]{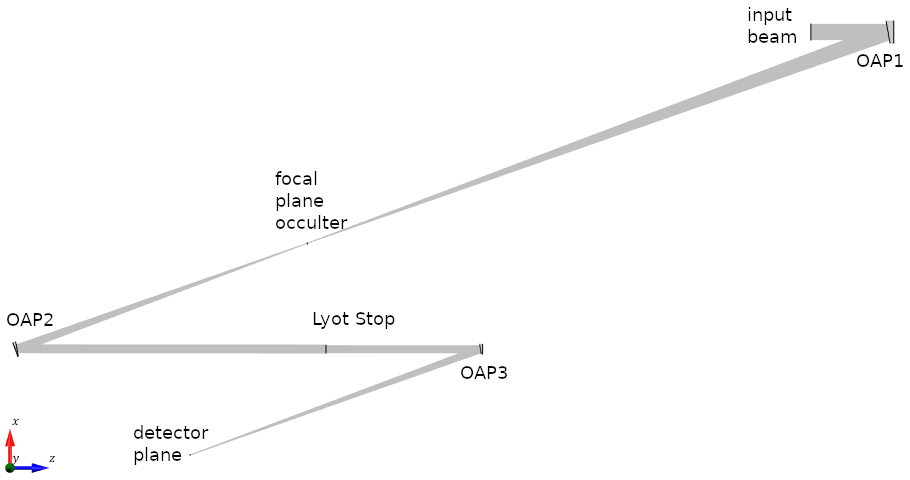}
	\caption{\small A schematic diagram of the Lyot-type stellar coronagraph used in these simulations. The DM, which is not shown, modulates the otherwise collimated input beam before OAP1.}
	\label{fig: coronagraph}
\end{figure}

\section{Simulations and Results}\label{sec: simulations}

All optical propagation computations were carried out using the VirtualLab Fusion (VLF) software from LightTrans, Inc.
The simulation results in this article correspond to a square Lyot-type stellar coronagraph shown schematically in Fig.~\ref{fig: coronagraph} and with key parameters summarized in Table~\ref{table: optical system}.
This coronagraph consists of 3 OAPs.
The initial propagation direction of the beam before it encounters OAP1 is in the $+z$ direction.
The simulation is configured so that the principal axis of the beam remains in the $x-z$ plane after each OAP encounter.
With this arrangement, the $y-$axis is invariant throughout the system.
Assuming no polarizers in the system, the consequence of this invariance is that $y-$polarized light in the input beam results in $y-$polarized light in the detector plane, weak polarization effects due to the finite $f$/\# not withstanding.
Similarly, if the light in the input beam is $x-$polarized it must also be $x-$polarized  in the detector plane since it must be orthogonal to the global $y-$ and local $z-$directions (again, weak polarization effects due to the finite $f$/\# not withstanding).   
Thus, the $A$ and $B$ directions discussed in  Sec.~\ref{sec: Intensity with CP} correspond directly to the $x$ and $y$ axes in the input beam.
To construct the $\tilde{\bD_x}$ and $\tilde{\bD_y}$ matrices introduced in Sec.~\ref{sec: models}, VLF carried out 1089 propagations-one for each DM actuator-through the optical system.  
For each of these propagations, the input beam was $x-$polarized and set to one of the DM influence functions with an amplitude of unity.
In the detector plane, the corresponding dominant field values, corresponding to a column of $\tilde{\bD_x}$, were the ``$E_x$'' values of the VLF output at each pixel, and the cross field values, corresponding to a column of $\tilde{\bD_y}$, were the ``$E_y$'' values.

A square coronagraph was chosen for convenience of equally illuminating all of the $33 \times 33$ DM actuators.
In Table~\ref{table: optical system}, the ``diameter,'' $D$ is the length of a side of the square, so,  the initial beam area is $D^2$.
Since the purpose of this article is to introduce several new probing and analysis concepts on a mostly realistic simulated system, not to model a specific laboratory system in detail, such conveniences are justified.

\begin{figure}[ht]
	\newcommand{\tw}{0.58\textwidth}
\vspace{-12mm}
\hspace{-2cm}
\begin{tabular}{r l}
	\includegraphics[width=\tw]{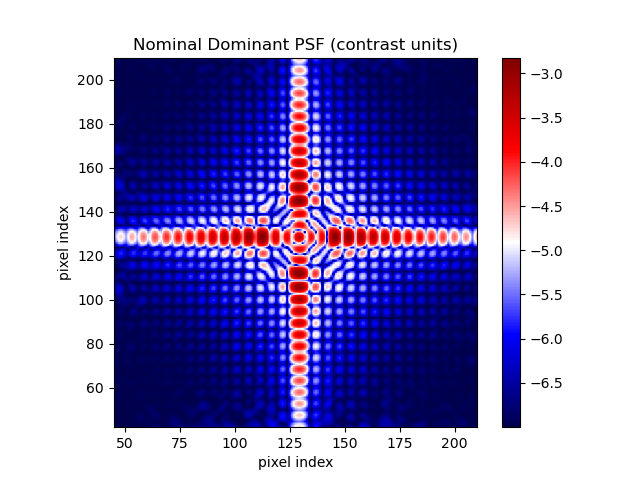} &
	\includegraphics[width=\tw]{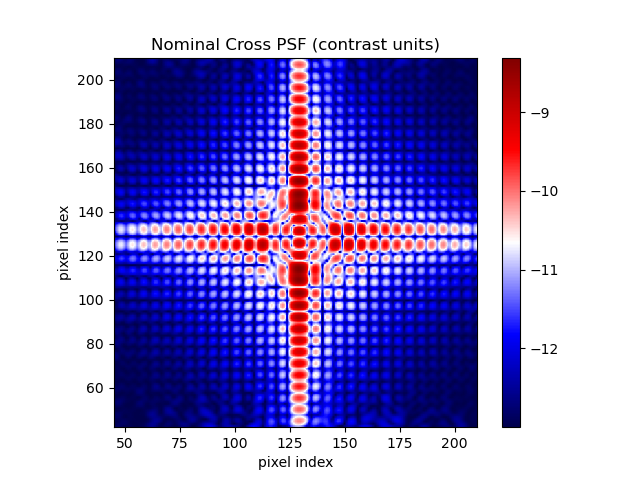} \\
	\includegraphics[width=\tw]{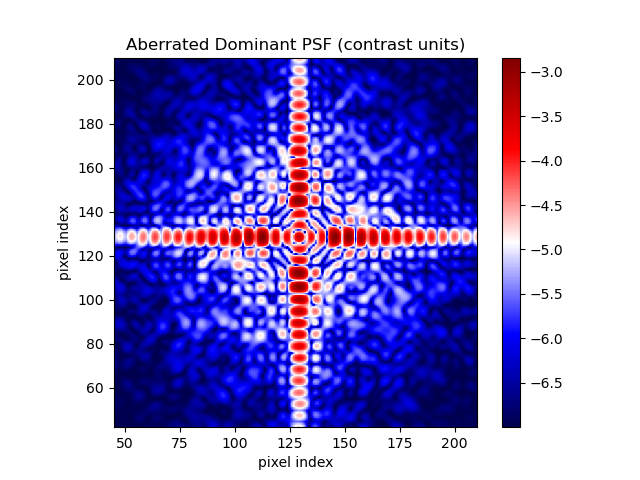} &
    \includegraphics[width=\tw]{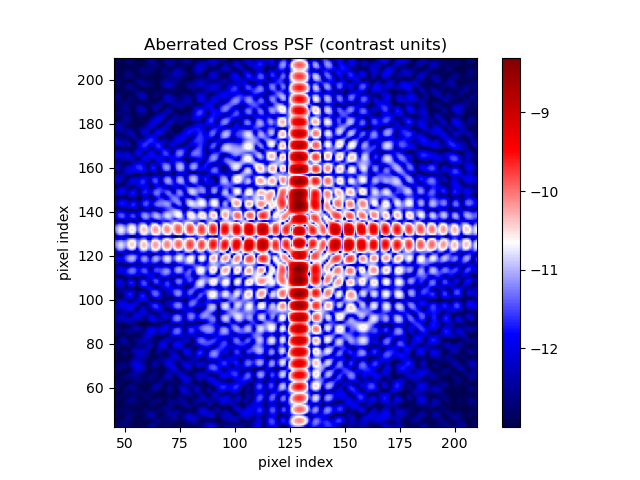} \\
	\includegraphics[width=\tw]{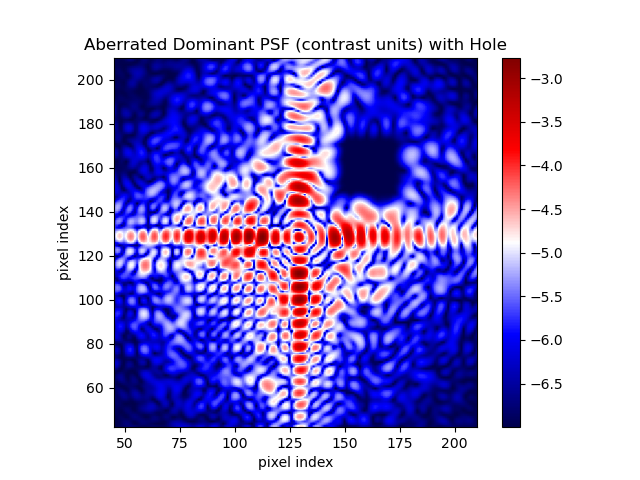} &
    \includegraphics[width=\tw]{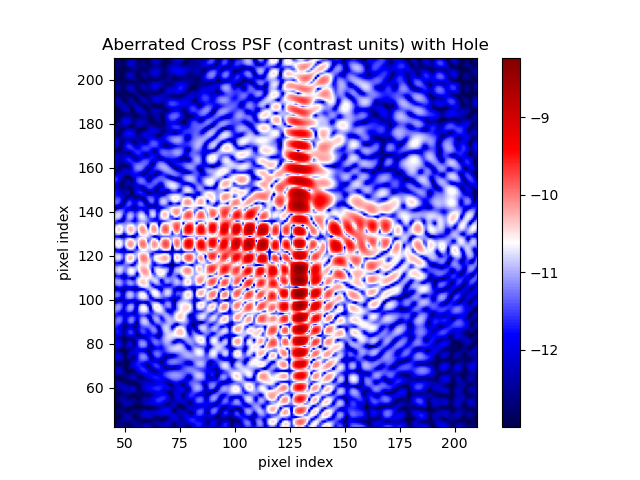} \\
\end{tabular}
\caption{\small PSFs from coronagraph simulations.  \emph{Left column:} Dominant intensity (contrast units).  
These values do not include the effect of the linear polarizers.
\emph{Right column:}  Cross intensity (contrast units).
\emph{Top Row:}  Nominal (i.e., aberration free) model.
\emph{Middle Row:}  Aberrated model.
\emph{Bottom Row:}  Aberrated model with a dark hole of size $4 \times 4 \, (\lambda/D)^2$.  
In the bottom-left image, the color scale obtains its minimum value at $10{-7}$ for display purposes.
Note the dark hole in the dominant intensity does not correspond to a dark hole in the cross intensity.}
\label{fig: PSFs}
\end{figure}

\begin{figure}[ht]
	\newcommand{\tw}{0.6\textwidth}
	\hspace{-1cm}
	\begin{tabular}{r l}
		\includegraphics[width=\tw]{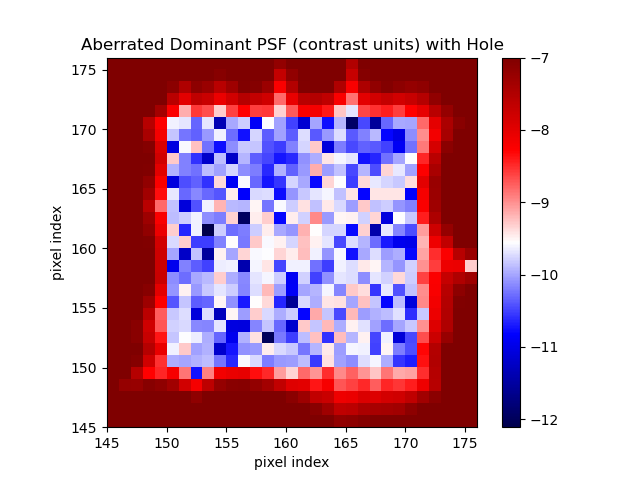} &
		\hspace{-16mm}
		\includegraphics[width=\tw]{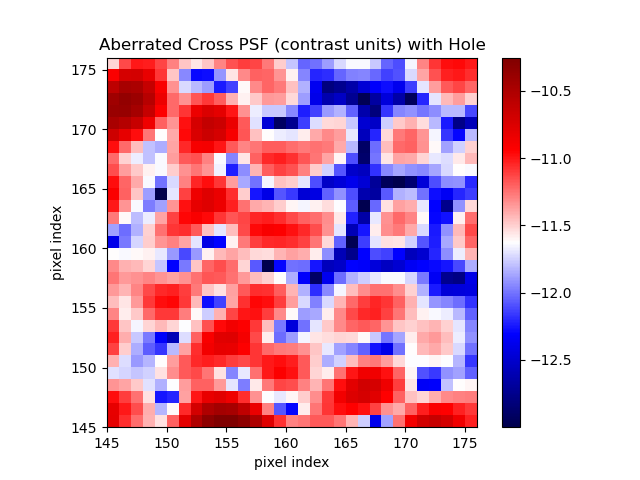}
	\end{tabular}
	\caption{\small Close-ups of the  $4 \times 4 \, (\lambda/D)^2$ dark hole region.
	\emph{Left:} The dominant intensity (contrast units).  \emph{Right:} The Cross intensity (contrast units). 
	These values do not include the effect of the linear polarizers.
 These are from the images shown in the bottom row of Fig.~\ref{fig: PSFs}. }
	\label{fig: HoleCloseUp}
\end{figure}

As explained in Sec.~\ref{sec: models}, the DM is modeled using cubic B-spline influence functions and unit-amplitude phasor coefficients.  
The DM, which is not shown in Fig.~\ref{fig: coronagraph} modifies the input beam, which is then propagated to OAP1 ($f = 0.8 \,$m).
Except for the DM modifications, the input beam is collimated.
OAP1 focuses the beam on the focal plane occulter, which is an opaque square $240 \, \mu$m ($5.8\, f \lambda/D$) on a side, with a $40 \, \mu$m wide border region over which the transmission smoothly goes from 0 to 1.  
OAP2 ($f = 0.4 \,$m) is placed $0.4 \,$m from the first focal plane and it places the Fourier transform of the field in OAP1's focal plane on the Lyot stop, which has a clear diameter of $9.1\,$mm and a transition border width of $1.2 \,$mm.   
Then, OAP3  ($f = 0.4 \,$m) makes the coronagraphic image in the detector plane.   
The aberration-free, or \emph{nominal}, PSF images corresponding to a plane wave input (which is simply a DM command of zero) for the dominant and cross intensities are provided in the top row of Fig.~\ref{fig: PSFs}.
Random phase and amplitude perturbations in the input beam created the aberrated images seen in the middle row of Fig.~\ref{fig: PSFs}.
The bottom row of Fig.~\ref{fig: PSFs} shows the aberrated PSFs with a dark hole in the dominant field with an area of $4  \times 4 \, (f \lambda/D)^2$.
This dark hole contains 441 detector pixels.
Given the known aberrated field in the detector plane from the simulation, determining a dark hole command was simply a matter of optimization.
Fig.~\ref{fig: HoleCloseUp} provides a close up of the dark hole region in the dominant and cross intensities, where it can be seen that a dark hole in the dominant intensity does not correspond to a dark hole in the cross intensity.

\begin{wraptable}{l}{0.55\textwidth}
	\begin{tabular}{|c|c|}
		\hline
		Parameter & Value \\
		\hline
		wavelength ($\lambda$) & $1 \, \mu$m \\
		input beam diameter ($D$) & 19.34 mm \\
		effective  $f$/\# & 44 \\
		DM actuators (spanning $D$) & $33 \times 33$ \\
		occulter opaque diameter & $240 \, \mu$m \\
		occulter transiton edge width & $40 \, \mu$m \\
		Lyot stop clear diameter & $9.1 \,$mm \\
		Lyot stop transition edge width & $1.2 \,$mm \\
		OAP1 focal length ($f$) & 800 mm \\
		OAP2, OAP3 focal length ($f$) & 400 mm \\
		OAP1, OAP2, OAP3 off-axis angle ($\zeta$) & $20^\circ$ \\
		\hline
	\end{tabular}
	\caption{\small Lyot coronagraph parameters used for the simulations.}
	\label{table: optical system}
\end{wraptable}
To perform the estimation of the cross fields in the dark hole, we take a contrast of unity to correspond to $10^{15}$ photons/pixel/exposure.
This is motivated by considering a common 100 mW continuous wave supercontinuum laser source that evenly distributes its power over a $1.6 \, \mu$m wavelength range.  
A 3\% bandpass filter, centered at the simulation wavelength of $1 \, \mu$m, allows about $10^{16}$ photons/s to pass through.  
The linear polarizer extinction coefficients $\epsilon_1$ and $\epsilon_2$ were both taken to be $2.5 \times 10^{-6}$ [see Eq.\eqref{eq: detector intensity w polarizers}], which is consistent with several off-the-shelf linear polarizers (also, sheet polarizers can be stacked).  
Fig.~\ref{fig: CF estimates} shows the estimates of the real (left panel) and imaginary (right panel) parts of the cross field in each of the 441 dark hole pixels.
The black squares are the true values and the red dots are the estimated values.
The error bars extending through the red dots are the square-roots of the error covariance matrices estimated by the Cram\'er-Rao bound, as explained in Sec.~\ref{sec: Regression}.

\section{Conclusions and Discussion}\label{sec: Conclusions}

\begin{figure}[]
	\newcommand{\tw}{0.53\textwidth}
\hspace{-12mm}
\begin{tabular}{r l}
	\includegraphics[width=\tw]{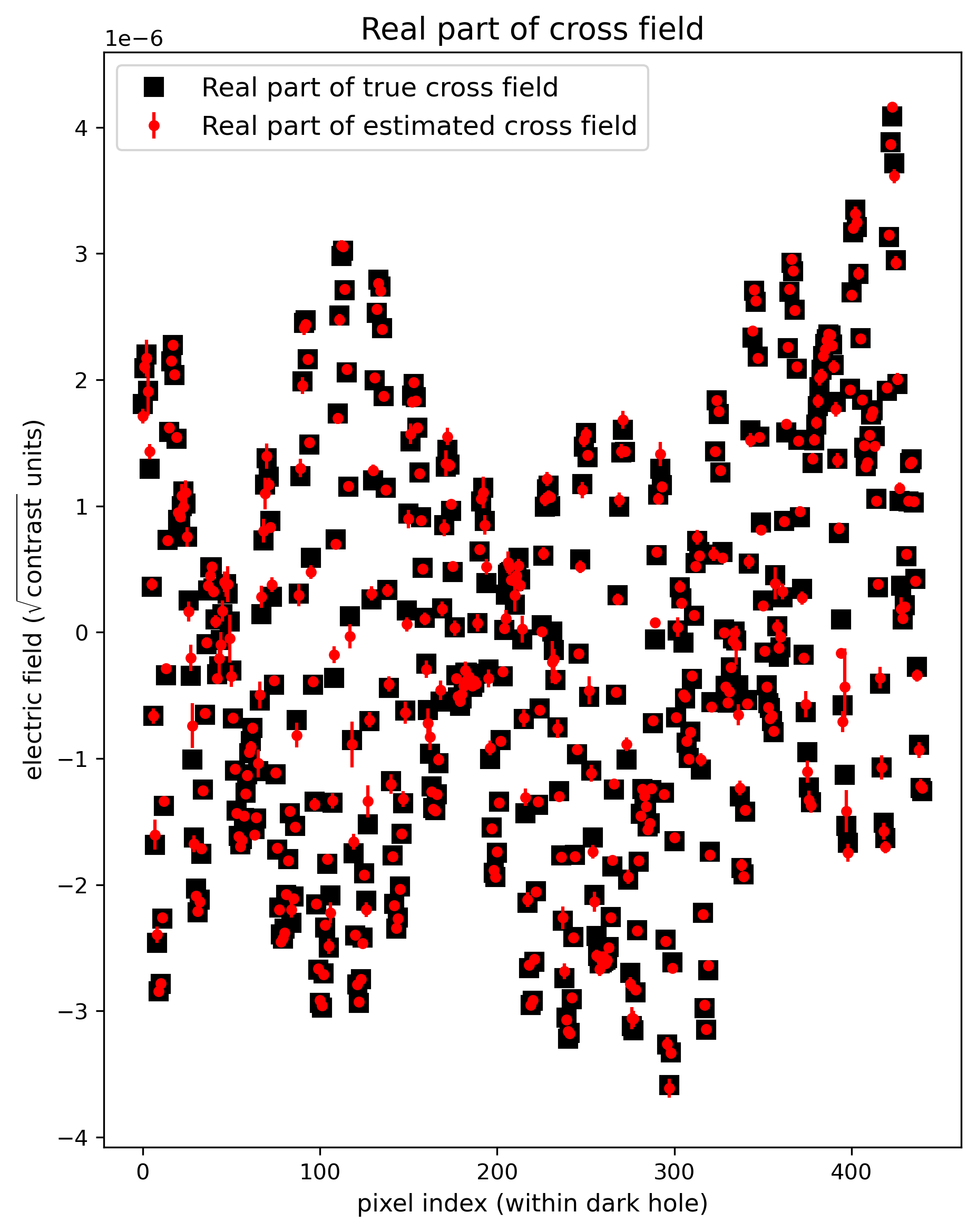} &
	\includegraphics[width=\tw]{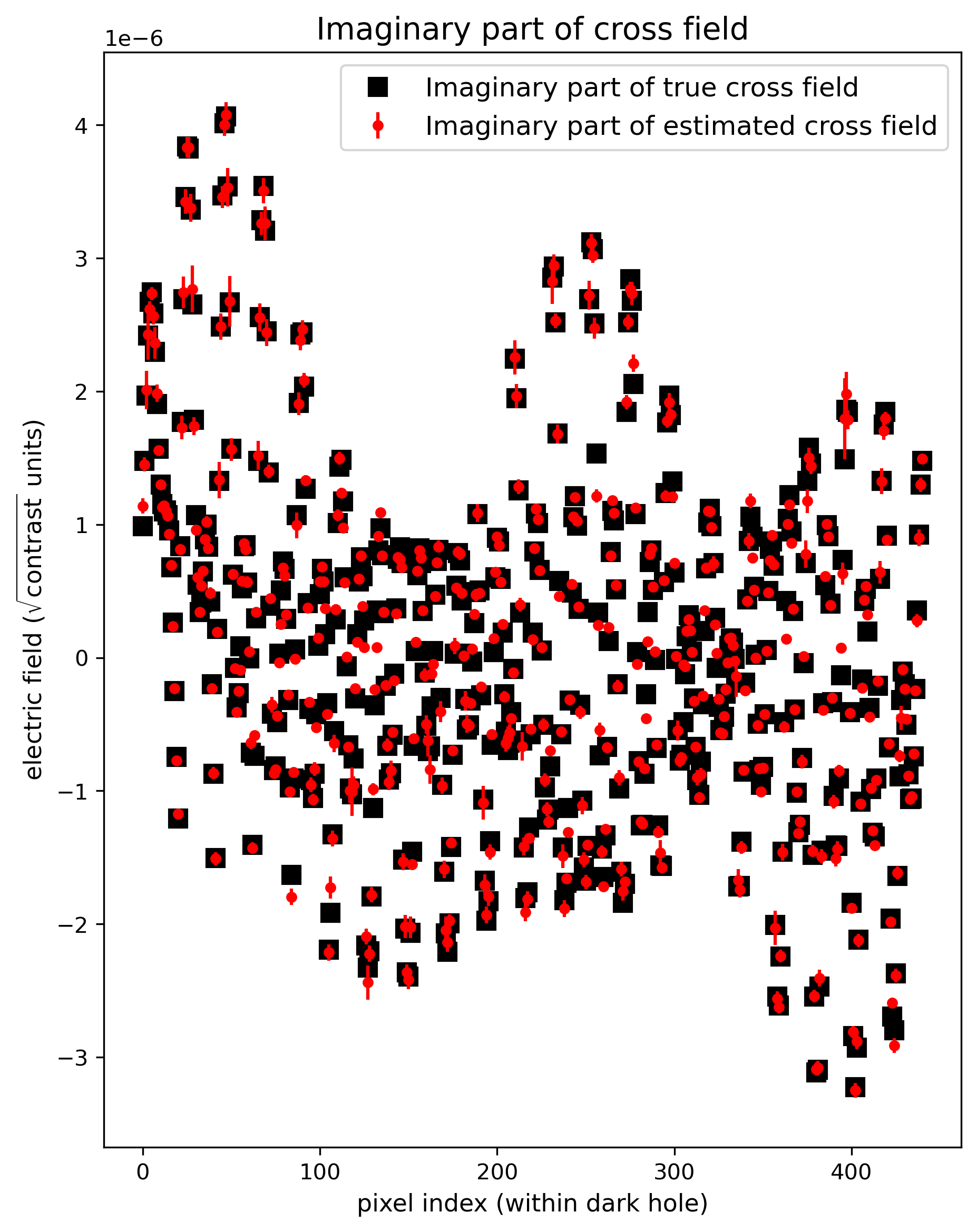}
\end{tabular}
	\caption{\small Estimated and truth values of the cross field in the dark hole region.
		\emph{Left:} Real part of the cross field.  \emph{Right:} Imaginary part of the cross field. 
		The black squares correspond to the true values of the fields and the red dots correspond to the probe-based nonlinear estimates described in the text.
		The error bars are square-roots of the variances estimated by the Cram\'er-Rao bound.}
		\label{fig: CF estimates}
\end{figure}

Cross-polarization effects are generally small, yet they are becoming more relevant as EFC progresses to ever greater contrast levels.
This article proposes and simulates a set of laboratory procedures for EFC-style probing measurements of the cross-polarization effects in coronagraphic direct imaging systems.  
Developing the capabilities to measure the cross-polarization in a laboratory setting is likely to be critical for validating models of the cross-polarization.
In these simulations, the probing and analysis procedures provided highly accurate estimates of the real and imaginary parts of the cross electric field within the dark hole, as shown in Fig.~\ref{fig: CF estimates}. 
Given the significant realism in these numerical simulations, these excellent results can be considered a demonstration of the viability of the proposed approach.  
Key elements of realism in the simulations include:
\begin{itemize}
\item The 3 OAP element Lyot coronagraph is designed with focal lengths and element sizes that are roughly similar to real testbed systems.
\item The random phase/amplitude mask applied in the entrance pupil performs a role that is similar to defects on various optical surfaces.
\item The simulation does not assume ideal linear polarizers in the sense that there is leakage of the orthogonal polarization into the beam.
\item No linear approximations are used in the various calculations.
\item The propagation of electric fields through the coronagraph is state-of-the-art and carried out with the VirtualLab fusion software.
\item The probe intensity measurements are from a Poisson random number generator to account for photon counting noise, which is the only significant noise source due the large number of counts from the laser.
\end{itemize}

Of course, the ultimate goal of an exoplanet mission is to measure the planetary light, represented as the incoherent component $\bh$ in Eq.~\eqref{eq: intensity probe X}.
On sky, regression equations similar to the ones presented in Sec.~\ref{sec: Regression}
 could be modified in order leverage model predictions of the cross field in order to make more accurate estimates of the dominant field and incoherent intensity.
Conveniently, the recent emphasis on digital twins provides opportunities for developing models of the cross field in the near future.\cite{Haffert_DigiTwin2024} 

A particularly useful form of model output would be a joint probability density function (PDF) of the real and imaginary parts of cross field, denoted as $\mathcal{P}$(cross field).
Access to $\mathcal{P}$(cross field) would allow formulation of a likelihood function of the form: \linebreak $\mathcal{P}$( measurements$\,|\,$incoherent intensity, dominant field, crossfield)$\times \mathcal{P}$(cross field), which could then be mariginalized over the cross field to yield a likelihood function of the form: \linebreak $\mathcal{P}$( measurements$\,|\,$incoherent intensity, dominant field).
The latter could be maximized to yield an estimate of the incoherent intensity that fully accounts for the cross polarization.
Although speculative, it is worth exploring the use of AI-based methods, such as convolutional neural networks, that predict the cross-polarization fields when the dominant field, likely measured with existing EFC methods, is provided as an input.
Mapping the dominant field to the cross field can be posed as an image translation problem, a problem well-suited to standard neural network (NN) architectures, such as the U-Net.\cite{UNET_ronneberger}
A digital twin of a coronagraph with random aberrations imposed could provide data sets for supervised learning to train such an NN.
Note that there are a variety of methods that employ NNs to provide estimates of posterior PDFs.\cite{barber2012bayesian}.

The proposed experiment and analysis method for measuring the cross-polarization electric field in the dark hole combines has the following key elements:
\begin{itemize}
	\item The use of linear polarizers to suppress the dominant polarization and enable measurement of the cross-polarization.
	\item The ability to self-consistently simulate the dominant and cross-polarization fields that arise due to various weakly polarizing effects in the optical system, including the finite $f$/\#. 
	\item Finding a set of candidate DM probes with the help of an optimization procedure.  The optimization procedure given here accounts for leakage in the linear polarizers.  This optimization procedure, in turn, is enabled by the next item:
	\item A method for rapidly calculating the image plane result of a DM command in both the dominant and cross fields, so that it can be carried out millions of times within a reasonable time frame (hours).  In this article, the matrix-based optical system model fulfilled this requirement.
	\item A procedure for evaluating the most informative set of DM probes to be used in the measurements from the above set of candidate choices. Here, this was done with an exhaustive-search combined a Cram\'er-Rao analysis.
	\item A procedure for effective solution of the nonlinear regression equations to make the final estimates.
\end{itemize}

The required laboratory set-up is likely to be reasonably straightforward on current high contrast imaging test beds, requiring only fairly modest equipment such as linear polarizers and a 100 mW continuous-wave supercontinuum laser.
In terms of carrying out a laboratory validation, a number of practical considerations come to mind:
\begin{itemize}
\item Laser fluctuations:  The power output and the polarization state of lasers often fluctuate in time.  Therefore, it is important to continually measure intensity of the beam after the first linear polarizer for the duration of the experiment.  One option for this that might be considered is measuring the intensity of the light back-reflecting from the focal plane occulter or mask.
\item Stray Light Contributions:  The light not transmitted or absorbed by the linear polarizers needs to be rendered harmless.  This may involve incarceration in light traps.
\item The linear polarizers will introduce aberration to the wavefront, causing the dark hole command, $\bc_0$ to change when the polarizers are inserted.
The dark hole command with both polarizers in place, $\bc_0'$ should obey the relationship: $\bc_0' \approx \bc_0 + \boldsymbol{\varsigma}_1 + \boldsymbol{\varsigma}_2$, in which  $\bc_0$ is the dark hole command without the polarizers, $\bc_0 + \boldsymbol{\varsigma}_1 $ is the dark hole command with only the first polarizer in place, and $\bc_0 + \boldsymbol{\varsigma}_2$ is the dark hole command with only the second polarizer in place.  If necessary, the procedures provided in this article can be generalized to jointly estimate the dominant field in addition to the cross field.
\end{itemize}

\clearpage

\subsection*{Disclosures}
The author has no conflicts of interest.

\subsection*{Data Availability Statement}
The Python codes and supporting data to reproduce the results given in this article are publicly available on GitHub at \emph{https://github.com/ComputationalAstrologer/Optics/tree/master/EFC}

\subsection*{Acknowledgments}
The author would like to acknowledge John Kohl, N. Jeremy Kasdin, Dan Sirbu and David Marx for helpful discussions.
This work was funded by the Heising-Simons Foundation (grant numbers: 2020-1826, 2022-3912) and 
the National Science Foundation (award number: 2308352).

\bibliographystyle{spiejour} 
\bibliography{CrossField.bib}
\end{document}